\documentclass[fleqn,usenatbib]{mnras}

\usepackage[T1]{fontenc}

\DeclareRobustCommand{\VAN}[3]{#2}
\let\VANthebibliography\thebibliography
\def\thebibliography{\DeclareRobustCommand{\VAN}[3]{##3}\VANthebibliography}

\usepackage{amsmath}
\usepackage{amssymb}
\usepackage{graphicx}
\usepackage{float}
\usepackage{parskip}
\usepackage{caption}
\usepackage{subcaption}
\usepackage{natbib}
\usepackage{newtxtext,newtxmath}

\title[Biosignatures for nutrient limited biospheres]{Predicting biosignatures for nutrient limited biospheres}

\author[A. E. Nicholson et al.]{
A. E. Nicholson,$^{1}$\thanks{E-mail: arwen.e.nicholson@gmail.com}
S. J. Daines,$^{2}$
N. J. Mayne,$^{1}$
J. K. Eager-Nash$^{1}$
T.M. Lenton$^{2}$
and K. Kohary$^{1}$
\\
$^{1}$Physics and Astronomy, College of Engineering, Mathematics and Physical Sciences, University of Exeter, Exeter, EX4 4QL, UK\\
$^{2}$Geography, College of Life and Environmental Sciences, University of Exeter, Exeter, EX4 4QD, UK
}

\date{This article has been accepted for publication in Monthly Notices of the Royal Astronomical Society Published by Oxford University Press on behalf of the Royal Astronomical Society.}

\pubyear{2022}

\begin{document}
\label{firstpage}
\pagerange{\pageref{firstpage}--\pageref{lastpage}}
\maketitle

\begin{abstract}
With the characterisations of potentially habitable planetary atmospheres on the horizon, the search for biosignatures is set to become a major area of research in the coming decades. To understand the atmospheric characteristics that might indicate alien life we must understand the abiotic characteristics of a planet and how life interacts with its environment. In the field of biogeochemistry, sophisticated models of life-environment coupled systems demonstrate that many assumptions specific to Earth-based life, e.g. specific ATP maintenance costs, are unnecessary to accurately model a biosphere. We explore a simple model of a single-species microbial biosphere that produces $CH_{4}$ as a byproduct of the microbes' energy extraction - known as a type I biosignature. We demonstrate that although significantly changing the biological parameters has a large impact on the biosphere's total population, such changes have only a minimal impact on the strength of the resulting biosignature, while the biosphere is limited by $H_{2}$ availability. We extend the model to include more accurate microbial energy harvesting and show that adjusting microbe parameters can lead to a regime change where the biosphere becomes limited by energy availability and no longer fully exploits the available $H_{2}$, impacting the strength of the resulting biosignature. We demonstrate that, for a nutrient limited biosphere, identifying the limiting nutrient, understanding the abiotic processes that control its abundance, and determining the biospheres ability to exploit it, are more fundamental for making type I biosignature predictions than the details of the population dynamics of the biosphere.
\newline
\end{abstract}




\section{Introduction}
\label{Section:introduction}

With recent instrumentation advances such as the launch of the James Webb Space Telescope and the Extremely Large Telescope (currently under construction), alongside future missions such as the Large Ultraviolet Optical Infrared Surveyor, searching for signs of life on planets beyond our solar system is set to be possible in the coming decades \citep{Snellen:2021, Quanz:2021}. 
The large diversity of exoplanets found to date indicate that potential biosignatures on different planets will likely manifest in different ways. Any potential biosignature must be understood within the context of its host planet \citep{Seager:2013a, Claudi:2017, Kiang:2018, Schwieterman:2018, Krissansen-Totton:2022}. One well known example of how planetary context is important for potential signs of life is the presence of atmospheric oxygen. The presence of oxygen in our atmosphere is a byproduct of biological processes and thus would act as a biosignature for remote observers of Earth, however high $O_{2}$ concentrations are possible abiotically for planets under different conditions to our planet \citep{Meadows:2018}. Finding and understanding any potential biosignatures will depend on our observational limits \citep{Fujii:2018}, our understanding of the abiotic processes at work on the candidate planet \citep{Catling:2018, Krissansen-Totton:2022}, and our understanding of how life interacts with its environment \citep{Lovelock:1974, Margulis:1974}.

Currently, Earth remains our only known example of a life--hosting planet and thus represents the natural starting point for understanding the possibility of detecting life elsewhere. However, Earth's biosphere has evolved significantly during its lifetime to date. Evidence for life on Earth, from the rock record, has now been found essentially at the earliest point possible \citep[e.g][]{Nisbet:2001} during the Archean period. The biosphere at the time was likely the simplest configuration and comprised of methanogens \citep{Schopf:2018}. Given that Earth spent roughly a third of its lifetime in the Archean \citep{catling2020archean}, it is natural to begin our study of the vast possibilities for biosignatures with this long--lived and comparatively simple biosphere. In this study we incorporate well studied principles from ecology and microbiology into a simple biosphere model, but allow the precise characteristics of our life to be free parameters, exploring the cases which ultimately support a stable population and, a potential, biosignature.

A framework for assessing potential biosignatures has been proposed by \cite{Catling:2018} where they suggest a probabilistic approach that combines observations of the candidate planet and its host star with models of the possible abiotic and biotic processes taking place on the planet to determine the probability of the planet being inhabited. Their framework proposes a process roughly following the outline: 1) characterising the stellar and exoplanetary system properties, including external exoplanet parameters (e.g. mass and size); 2) characterising of internal exoplanet properties (e.g. climate); 3) assessing potential biosignatures within the environmental context; and 4) exclusion of false positives. Only by assessing all these details can we make predictions as to whether the presence of a certain feature in a planet's atmosphere is likely due to life or abiotic processes.

This work falls under step 3) of the \cite{Catling:2018} process. Understanding a potential biosignature for any particular planet requires models to help us understand what processes we expect to be happening on the planet in the absence of life, and how life would interact with its planet. While determining the potential metabolic pathways for life will be vital when considering possible biosignatures, we will demonstrate that for a simple model biosphere limited by nutrient availability understanding the underlying population dynamics of the biosphere is not necessary to predict the `strength' of the biosignature produced. Instead understanding the limiting nutrient is more fundamental and, as we will demonstrate, large differences in the total population of the biosphere only result in small differences in the strength of a biosignature. Population dynamics in this work refers to factors such as the total population of the biosphere, and the rates of death and reproduction, but not to any motion of microbes moving in their environment.

Although this insensitivity of the population dynamics of life to its larger scale impact, and the importance of limiting factors, is well known across studies of Earth history \citep{herman:2005, Kharecha:2005, bruggeman:2014, Lenton:2018, zakem:2020}, its implications represent an important shift in our approach to biosignatures. Models of biogeochemistry used to investigate Earth's climatic history, such as the Archean environment \citep{Kharecha:2005} or the rise of oxygen in Earth's atmosphere \citep{Lenton:2018}, use sparse data from Earth history to recreate past climates, and much of this research has implications for the search for biosignatures. With our planet as the only known home to life, our assumptions about possible alien life will be biased by the life, both past and present, that we find on Earth. However if we can minimise the number of assumptions needed to model alien life, and avoid as many Earth-life assumptions as possible, we can formulate robust predictions for how potential biosignatures might manifest on alien worlds. In this work we take a step towards this goal. Using a simple model, built on fundamental principles of microbial life on Earth, we demonstrate that for a nutrient limited single-species biosphere the ability of the biosphere to exploit its limiting nutrient is more fundamental to determining the planet's biosignature than the total population. Future work (Daines et al. in Prep) aims to expand on this goal towards a minimal model of biology for more general use in forming biosignature predictions.

A classification of gaseous biosignatures has been proposed by \cite{Seager2013} is as follows: type I biosignatures are generated as a by-product from microbial energy extraction. Type II biosignatures are gases produced as byproducts from building biomass, and type III biosignatures are those which are produced by life but not as by-products of their central chemical functions. In this work we consider type I biosignatures in the above classification scheme and any references to biosignatures in this work will refer to this classification of biosignature unless otherwise stated.

We present a highly simplified model of an Archean-Earth-like planet, home to a single species of life which produces methane as a byproduct of energy extraction. On our model planet there is no abiotic source of methane allowing us to take this gas as a clear biosignature. We demonstrate that, assuming the microbes ability to exploit the limiting resource (in this case $H_{2}$) remains unchanged, the details of the population dynamics of the biosphere are largely irrelevant to the abundance of methane in the atmosphere. Instead we demonstrate that the availability of the limiting resource, in this case hydrogen, has a much stronger impact on the abundance of atmospheric methane. In this study we model the abiotic environment and the microbe behaviour in a highly simplified manner to allow us to determine the relationship between the population dynamics of the biosphere and the resulting biosignature more easily. This study acts a step forward in complexity from more abstract models of life-environment coupled systems \citep{williams2007flask, Nicholson:2017, nicholson2018alternative, Nicholson:2018, alcabes:2020} inspired by realistic models of biogeochemistry designed to recreate ecosystems and climates in Earth history \citep{herman:2005, Kharecha:2005, bruggeman:2014, Lenton:2018}. 

This work takes a step towards the goal of determining the minimal biological assumptions needed to contemplate possible type I biosignatures in scenarios where life is limited by nutrient availability.
Future work will certainly be necessary to build on the complexity of the present approach, and enhance the range and precision of the parameters controlling the various processes. Our goal here is just to focus on the simplest problem possible to begin the journey of understanding the key elements of the wider life-planet interaction.

The paper is structured as follows, in Section \ref{Section:modelsetup} we detail our simple model setup for both the planet and the microbes. In Section \ref{Section:experimentsetup} we outline the specific experiments we have performed, before presenting and discussing our model results in Section \ref{Section:H2_limited}. Finally, we extend our work to include more realistic process for the amount of energy generated by the microbe's metabolisms in Section \ref{Section:thermoresults}, before concluding in Section \ref{Section:summary} and looking forward for future potential steps in Section \ref{Section:nextsteps}.

\section{Model setup}
\label{Section:modelsetup}

We simulate a highly simplified zero-dimensional Archean Earth-like planet covered in a global ocean. We keep the model setup simple in order to develop a tractable framework for exploring the interactions between population dynamics and the atmosphere for a simple methanogen biosphere living in an Archean Earth-like environment.

Life on Earth emerged at least 3.8 billion years ago \citep{Woese:1977, Nisbet:2001} and methanogens - life that consumes $H_{2}$ and $CO_{2}$ and excretes $CH_{4}$ are thought to be some of the earliest lifeforms \citep{Schopf:2018}. We base our model on a planet with newly emerged life before huge diversification occurred. In our model we consider only a single species of life - single-celled methanogens. We restrict life to the ocean of our planet, and assume that both the ocean and atmosphere are well-mixed. This simplification is justified as we are not modeling the atmosphere or ocean over short timescales. We explore scenarios where microbe growth is limited by the availability of $H_{2}$ to the ocean. We assume the microbe uptake of $H_{2}$ is limited only by availability, and so the biosphere is able to fully exhaust $H_{2}$ in the ocean. This is a simplification on how nutrient uptake occurs in real microbes (see Section \ref{Section:microbes} for further discussion).

\begin{figure}
\centering
\begin{subfigure}{.48\textwidth}
  \centering
  \includegraphics[scale=0.3]{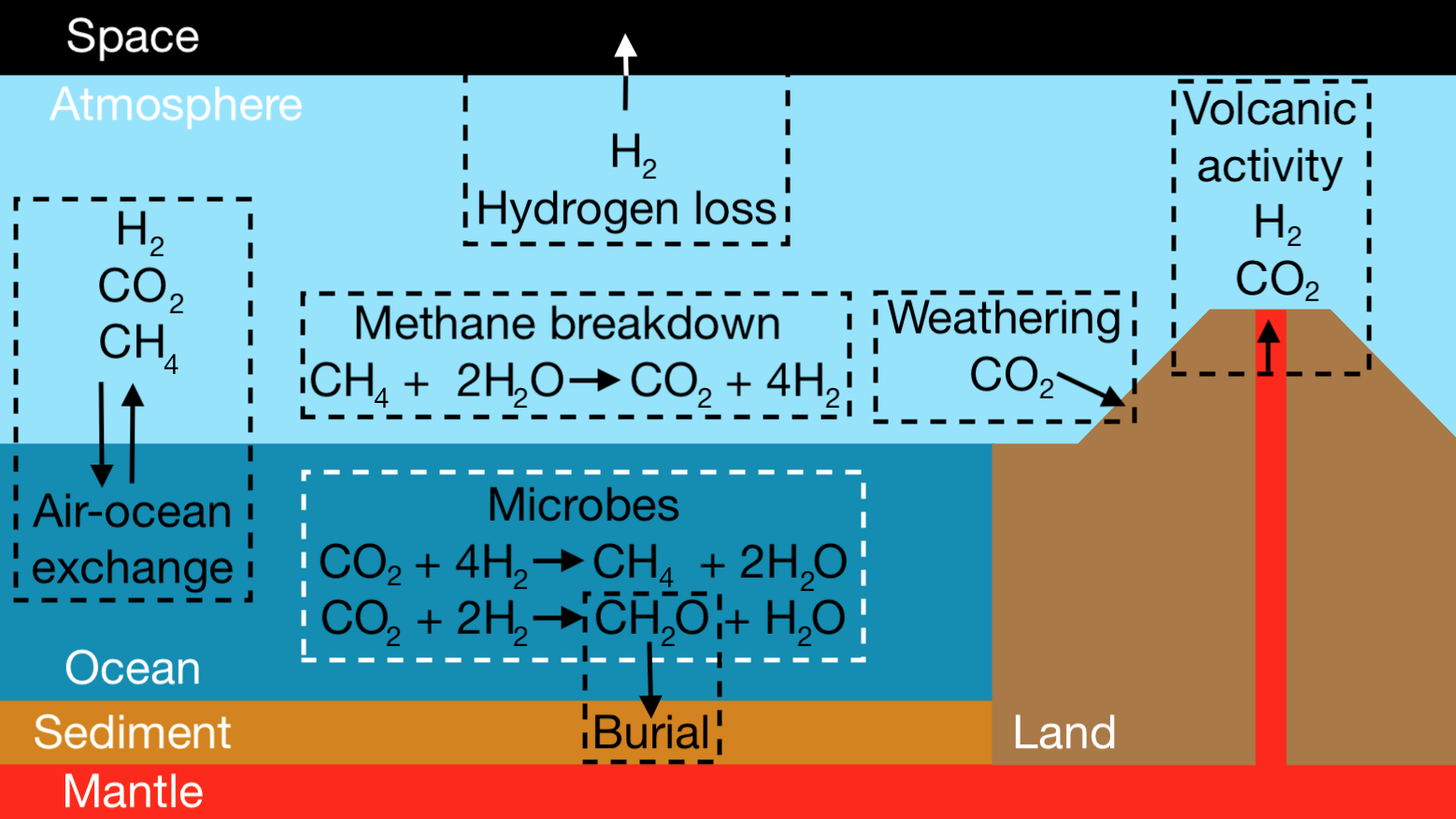}
\end{subfigure}%
\caption{A schematic showing the key abitoic (black dashed boxes) and biotic (white dashed box) processes occurring in our model.}
\label{fig:1}
\end{figure}

In order to capture the primary mechanisms determining the interaction of life with a planetary climate we need to describe the cycling of the key chemicals in the system, the response of the planetary climate to changes in composition, and the processes performed by or controlling life (such as cycling of chemical species, population growth, death, etc.), here described as a population of single-celled organisms. In the following sections we describe how we capture each of these key elements as simply as possible. Figure \ref{fig:1} shows a schematic of our model demonstrating the key processes occurring on the planet. Abiotic processes are shown surrounded by a black dashed box, and biotic processes by a white dashed box.

\subsection{Planet setup}

We track three chemical species in our planet's atmosphere and ocean: $H_{2}$, $CO_{2}$ and $CH_{4}$. The rest of the atmosphere is assumed to be made up of $N_{2}$ as this is the most abundant part of Earth's atmosphere. We assume abundant $H_{2}O$ is available for all chemical reactions requiring it. We also assume the atmospheric pressure ($P_{atmo}$) and total number of moles of gas in the atmosphere ($n_{atmo}$) remain constant throughout each experiment, at $P_{atmo} = 1$ atm, and $n_{atmo} = 1.73\times 10^{20}$ moles, taking modern Earth values. We update the abiotic environment in our model in timesteps representing years.

\subsubsection{Atmospheric $H_{2}$}

We assume a source of $H_{2}$ via an approximation of outgassing from volcanoes. We also assume a constant rate of removal of $H_{2}$ from the atmosphere, although this processes is different to the way $CO_{2}$ is removed. On Earth $H_{2}$ is irreversibly lost to space via hydrogen diffusion out of the upper layers of our atmosphere. The rate of this loss depends on the mixing ratios of the hydrogen bearing chemical species in the stratosphere \citep{Hunten:1973, Walker:1977}. If we assume a dry stratosphere, the rate of hydrogen loss is proportional to $f(H_{2}) + 2\ f(CH_{4})$ where $f(H_{2})$ and $f(CH_{4})$ are the mixing ratios of $H_{2}$ and $CH_{4}$. To keep with the simplified nature of our model, to perform $H_{2}$ loss from our model atmosphere we remove a percentage of $H_{2}$ proportional to $T(H_{2}) + 2T(CH_{4})$ where $T(H_{2})$ and $T(CH_{4})$ are the total number of moles of $H_{2}$ and $CH_{4}$ in the atmosphere.

\subsubsection{Atmospheric $CO_{2}$}

We assume a constant source of $CO_{2}$ to the planet's atmosphere in an approximation of volcanic outgassing. $CO_{2}$ is removed by removing a fixed percentage of the atmospheric $CO_{2}$ each year. This is a huge simplification of silicate weathering - a chemical process that removes $CO_{2}$ from Earth's atmosphere that is temperature and humidity dependant \citep{Brady:1994}. During weathering $CO_{2}$ reacts with minerals in surface rocks and is removed from the atmosphere. In our model we set the abiotic influx and rate of outflux of $CO_{2}$ in the atmosphere to be kept constant for the duration of each experiment. As we will be exploring scenarios where $H_{2}$ is the limited resource on microbe growth the details of the $H_{2}$ cycle on our planet are more important to microbe metabolic activity than those of the $CO_{2}$ cycle allowing us to use a simplified mechanism for $CO_{2}$ removal from the atmosphere. In addition we are not exploring timescales long enough for impacts of a brightening star to have an effect on the planet's climate.

\subsubsection{Atmospheric $CH_{4}$}

In our model there is no abiotic source of $CH_{4}$, however as we will detail in Section \ref{Section:microbes} our model microbes excrete $CH_{4}$ as a byproduct of their metabolism. Avoiding any abiotic source of methane makes measuring the hypothetical biosignature on the planet easier as any methane in the atmosphere must be due to biological activity. Additionally, biotic production of methane was much higher than abiotic production during the Archean \citep{KASTING:2005} the period of Earth history our model is loosely based on. Today, $CH_{4}$ is rapidly oxidised limiting the buildup of methane in the atmosphere, but in the low $O_{2}$ atmosphere of the Archean, $O_{2}$ would have rapidly been rapidly consumed and $CH_{4}$ was long lived \citep{Catling:2001}. On real planets the breakdown of atmospheric methane by photolysis is a complex process that depends on the altitude of the $CH_{4}$ and involves several stages. Here we simplify this and assume that methane breaks down back to $CO_{2}$ and $H_{2}$ in the atmosphere following

\begin{equation}
\label{Equation:methanbreakdown}
    CH_{4} + 2H_{2}O \rightarrow CO_{2} + 4H_{2}.
\end{equation}

This reaction is a combination of the overall methane photolysis pathway of methane at high altitudes following 

\begin{equation}
\label{Equation:methanephotolysis}
    CH_{4} + 2CO_{2} \rightarrow 2CO + 4H,
\end{equation}

and $CO$ removal performed by life a via a few metabolisms that combine to form

\begin{equation}
\label{Equation:cotoch4}
    4CO + 2H_{2}O \rightarrow 3CO_{2} + CH_{4},
\end{equation}

following the work of \cite{Kharecha:2005}. Equations \ref{Equation:methanephotolysis} and \ref{Equation:cotoch4} can then be combined to produce Equation \ref{Equation:methanbreakdown}, our process for $CH_{4}$ removal from the atmosphere. Without a process removing $CO$ it will rapidly build up in the atmosphere, known as CO runaway \citep{Zahnle:1986, Kasting:1983}, and become an abundant energy resource for life to exploit. We therefore assume a process removing $CO$ from the atmosphere without modeling an additional life form in our model. The experiments presented in this paper will be limited by $H_{2}$ availability, not $CO_{2}$, and so it is sufficient to say that the carbon cycle is closed without too much concern as to the nature of the process removing $CO$ from the atmosphere. If however our experiments were carbon limited, this simplification would break down as the details of the carbon cycle would become more important than those of the hydrogen cycle. The availability of $CO$ to any $CO$ consuming life-form will be dependent on the metabolic waste of methanogens - the life form excreting the $CH_{4}$ that in turn is photolysised to produce $CO$ (Equation \ref{Equation:methanephotolysis}). Therefore methanogens are the primary producers for the ecosystem and their population dynamics will determine the dynamics of lifeforms reliant on their waste products. We assume That Equation \ref{Equation:methanbreakdown} occurs at a constant slow rate.

These abiotic processes are simplifications of much more complex processes that occur on planets. On a real planet these processes will change over time and depend on many factors, e.g. changing tectonic activity or involving temperature dependence. As we are interested in the overall behaviour of the simple life-environment coupled system and are not trying to recreate the climate of a real planet we use these simplifications to keep the abiotic environment simple while tracking the abundances of $CH_{4}$, $CO_{2}$ and $H_{2}$.

Table \ref{table1} shows the values for the influxes and outfluxes of $CO_{2}$, $H_{2}$ and $CH_{4}$ for our system. These are kept fixed throughout our experiments. For methane, the outflux is the percentage of atmospheric $CH_{4}$ that undergoes the process described by Equation \ref{Equation:methanbreakdown} per year.

\begin{table}
\caption{Parameters for the influx and outflux of atmospheric $CO_{2}$, $H_{2}$ and $CH_{4}$ where $T(X)$ is the total number of moles of molecule $X$ in the atmosphere.}
\begin{center}
\begin{tabular}{|c|c|c|} \hline
Chemical & Influx (year$^{-1}$) & Outflux (year$^{-1}$)\\ \hline \hline
$CO_{2}$ & $10^{16}$ & $0.001\times T(CO_{2})$  \\ \hline
$H_{2}$ & $10^{14}$ &  $0.001\times (T(H_{2})+2T(CH_{4}))$ \\ \hline
$CH_{4}$ & 0 & $0.001\times T(CH_{4})$  \\ \hline
\end{tabular}
\end{center}
\label{table1}
\end{table}%

\subsubsection{Ocean-atmosphere gas exchange}

\begin{figure}
\begin{subfigure}{.48\textwidth}
  \centering
  \includegraphics[scale=0.45]{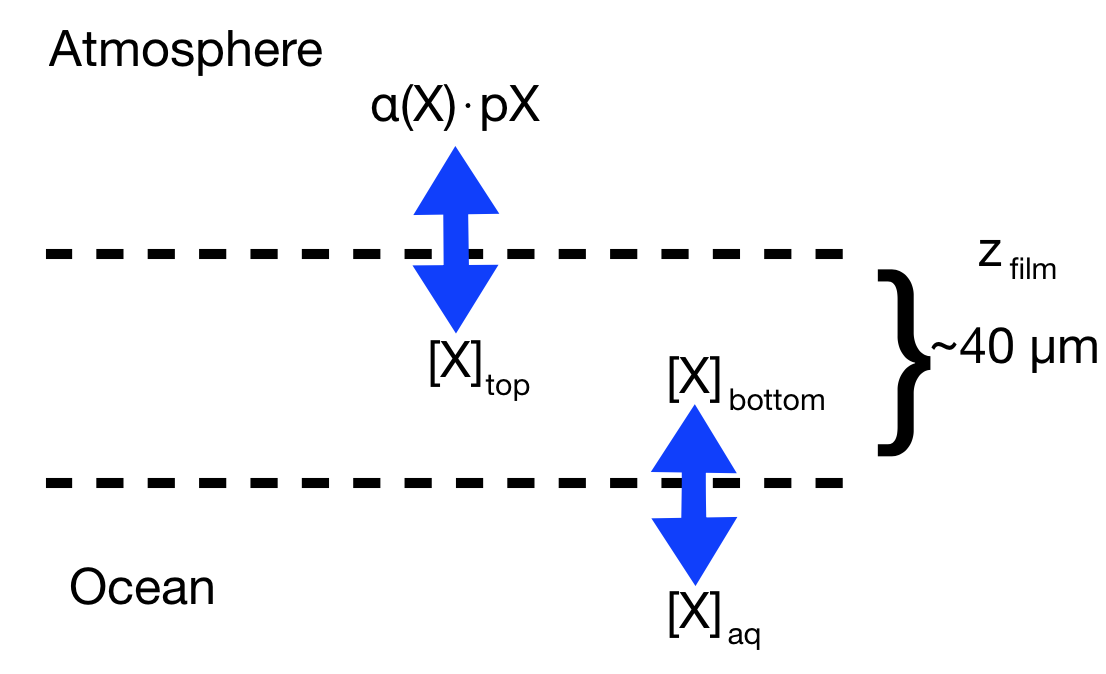}
\end{subfigure}
  \caption{The stagnant layer model for gas exchange between the atmosphere and ocean through a thin film of thickness $z_{film}$, where $\alpha(X)$ solubility of $X$ (i.e. the Henry's law coefficient), $pX$ the partial pressure of $X$ (in Bar), $ \left[ X \right]_{aq}$ is the dissolved concentration of $X$ (in $mol\ m^{-3}$), and $ \left[ X \right]_{top}$ $ \left[ X \right]_{bottom}$ are the concentrations of $X$ at the top and bottom of the film respectively.}
\label{fig:stagnant_layer}
\end{figure}

The gases in the model atmosphere can dissolve into the global ocean where they become available to life. In another simplification we assume no other source of $H_{2}$ or $CO_{2}$ to the ocean except that which dissolves into the ocean, and assume no outflux other than outgassing into the atmosphere. We calculate the transfer of gas between the atmosphere and ocean following the stagnant boundary layer model \citep{Liss:1974}. We assume the rate of exchange of gases between the atmosphere and the ocean depends on the concentration gradient of those gasses through a very thin film on the top of the ocean - the stagnant layer. Figure \ref{fig:stagnant_layer} shows a schematic of this model for how gases are exchanged between the ocean and atmosphere. 

The rate of exchange of a chemical between the atmosphere and ocean is set by its relative concentrations

\begin{equation}
\label{Equation:airseaexchange}
\Phi_{X} = v_{p}(X) \cdot (\alpha(X) \cdot pX - \left[ X \right]_{aq}),
\end{equation}

where $\Phi_{X}$ is the molecular flux of a chemical species $X$ from the atmosphere into the ocean, 
$v_{p}(X)$ is the piston velocity of the chemical species $X$ (which can be thought of as the speed at which a gas is being pushed into (or out of) the water column), 
$\alpha(X)$ solubility of $X$ (i.e. the Henry's law coefficient), 
$pX$ partial pressure of $X$ (in Bar), and 
$ \left[ X \right]_{aq}$ is the dissolved concentration of $X$ (in $mol\ m^{-3}$). We calculate the dissolved concentration of $H_{2}$, $CO_{2}$ and $CH_{4}$ in the ocean assuming an ocean depth of 100m. $v_{p}(X)$  is calculated by dividing the thermal diffusivity of $X$ by the thickness of the film $z_{film}$, which we assume to be to be $z_{film} = 40 \mu m$ \citep{Kharecha:2005}. Table \ref{table2} shows the values for the parameters needed to calculate the gas exchange between the atmosphere and the ocean for $CO_{2}$, $H_{2}$ and $CH_{4}$.

\begin{table*}
\caption{Parameters for the atmosphere-ocean exchange of $CO_{2}$, $H_{2}$ and $CH_{4}$. $*$: $CH_{4}$ and $H_{2}$ values from \protect\cite{Kharecha:2005}, $\alpha$: $CO_{2}$ diffusivity value from \protect\cite{Zhang:2018}, $\beta$: $CO_{2}$ solubility from \url{https://webbook.nist.gov/chemistry/}. Piston velocities are calculated assuming a stagnant boundary layer thickness of $z_{film} = 40 \mu m$. We assume 25$^{o}$C for the values for our gases and keep this fixed.}
\begin{center}
\begin{tabular}{|c|c|c|c|} \hline
Chemical  & Piston velocity ($m\ s^{-1}$) & Diffusivity ($m^{2}s^{-1}$) & Solubility $mol\ L^{-1}bar^{-1}$ \\ \hline \hline
$CO_{2}$ & $6.7\times 10^{-4}$ &  $2.67\times 10^{-6}$ $^\alpha$ &  $0.035$  $^\beta$ \\ \hline
$H_{2}$  & $1.3\times 10^{-2}$ $^*$ & $5.0\times 10^{-5}$ $^*$ & $7.8\times 10^{-4}$ $^*$ \\ \hline
$CH_{4}$ & $4.5\times 10^{-3}$ $^*$ & $1.8\times 10^{-5}$ $^*$ & $1.4\times 10^{-3}$ $^*$  \\ \hline
\end{tabular}
\end{center}
\label{table2}
\end{table*}

\subsubsection{Temperature dependence on $CH_4$ and $CO_2$}
\label{Section:temp_CH4_CO2}

The surface temperature of a planet, in the absence of life, will depend on many properties of the planet and host star. Furthermore, several biotic processes can impact the planetary climate. For our simplified system we capture the life-climate interaction through a parameterised treatment of the surface temperature as a function of atmospheric composition, generated using an idealised general circulation model. Our model life then changes the planet's atmospheric composition via its metabolic activity (Section \ref{Section:microbes}) and in turn impacts the average surface temperature. We use the Met Office Unified Model (UM) - a climate model adapted for exoplanets \citep{Boutle:2017, Eager:2020} to capture `snapshots' of the planetary temperature for differing atmospheric $CO_{2}$ and $CH_{4}$ concentrations. Both these important greenhouses gases are thought to have been more abundant in the Archean when life emerged and have provided significant warming \citep{Catling:2020}, so we restrict ourselves to considering the temperature dependence on only these two gases. In this study we will focus on methane as this is the atmospheric biosignature produced by the microbes on our model planet.

To generate our $CO_{2}$, $CH_{4}$ and temperature relationship we run a UM simulation setup as described in \cite{Eager:2022}. This is essentially a version of the Global Atmosphere configuration 7.0 \citep{Walters:2019} adapted to the Archean Earth, with a simplified slab ocean and surface with constant radiative properties. The bulk properties defining a planetary system within the UM - the planet radius, and the properties of the host star, are given in Table \ref{Table:0}.

\begin{table}
\caption{Bulk parameters for the zero-dimensional Archean Earth-like model planet.}
\begin{center}
\begin{tabular}{|c|c|} \hline
Parameter & Value \\ \hline \hline
Planet radius & $6051.3 \times 10^{3} m$ \\ \hline
Star spectral class & G-type main sequence \\ \hline
Star age & 1 billion years \\ \hline
\end{tabular}
\end{center}
\label{Table:0}
\end{table}%

\begin{figure}
\centering
\begin{subfigure}{.48\textwidth}
  \centering
  \includegraphics[scale=0.48]{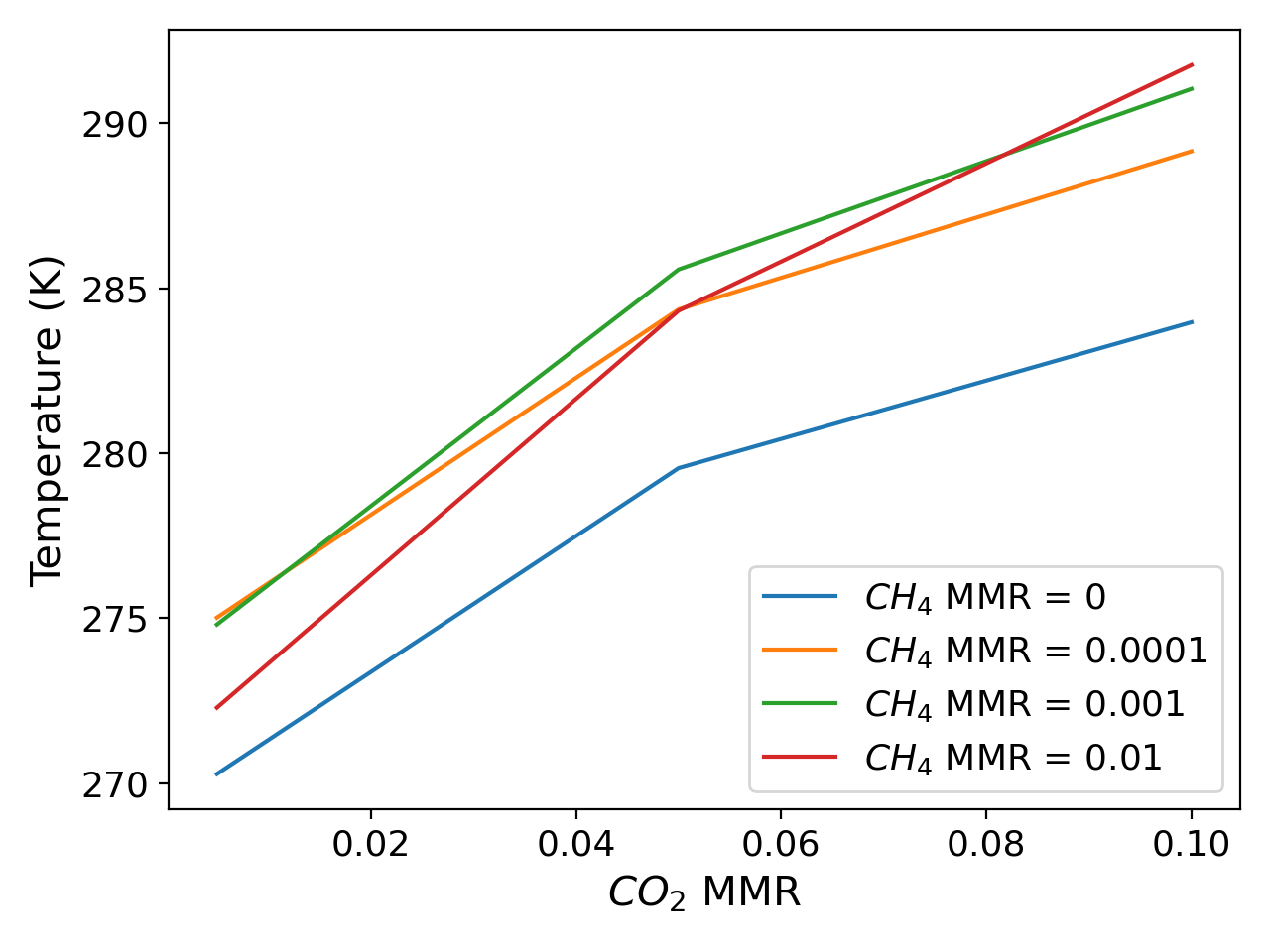}
  \caption{`Snapshots' obtained from the Unified Model.}
  \label{fig:sub2a}
\end{subfigure}%

\begin{subfigure}{.48\textwidth}
  \centering
  \includegraphics[scale=0.50]{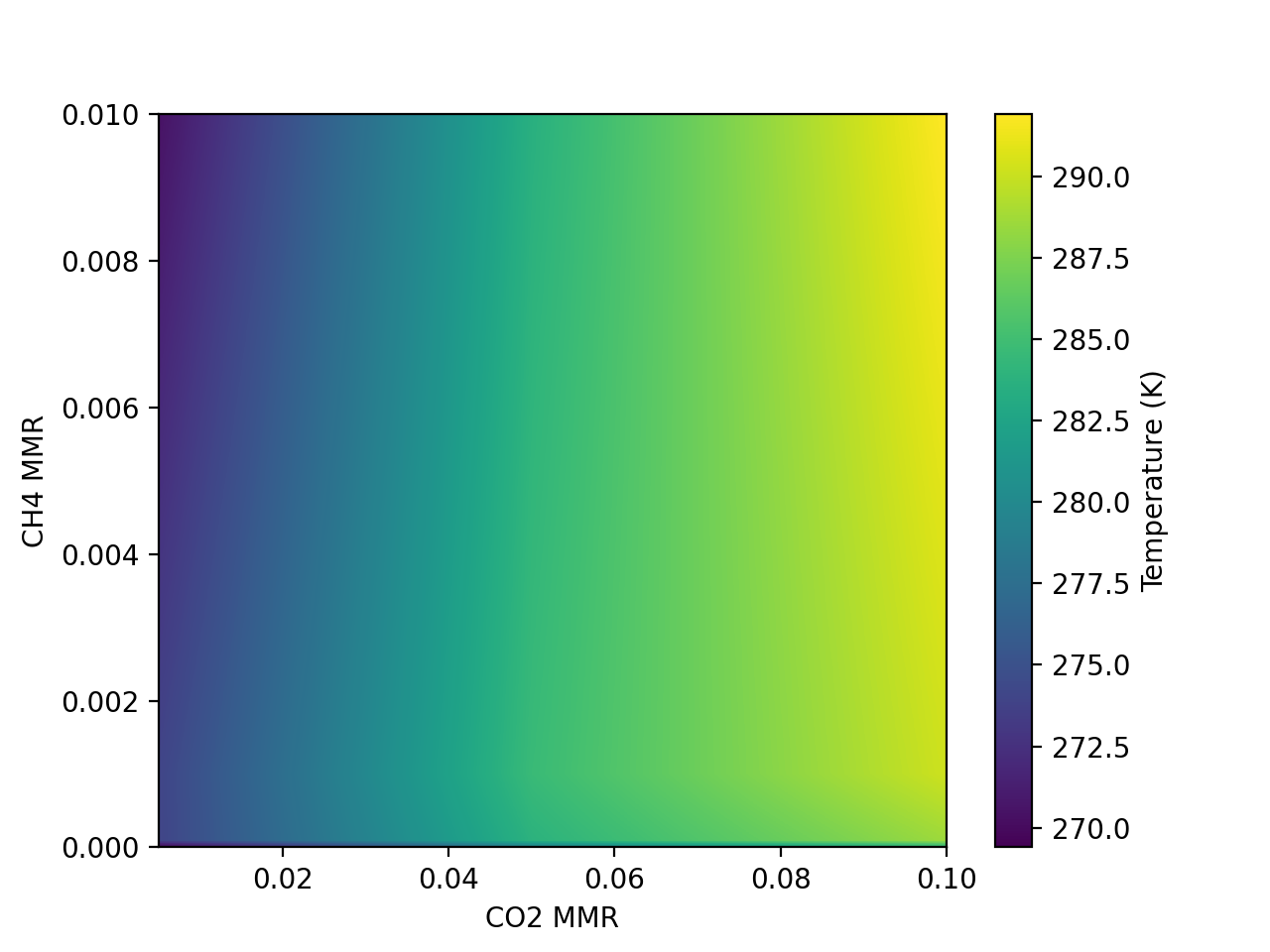}
  \caption{Interpolated data.}
  \label{fig:sub2b}
\end{subfigure}
\caption{Global average surface temperature dependence on atmospheric levels of $CH_{4}$ and $CO_{2}$ mass mixing ratios.}
\label{fig:2}
\end{figure}

To generate data points for the average surface temperature dependence on atmospheric $CO_{2}$ and $CH_{4}$, we run a UM configuration for the desired atmospheric composition for 10 years to reach equilibrium and then run for another 10 years and average the surface temperature for this time. Within the UM the atmospheric abundance of a gas is given in terms of the mass mixing ratio - this is the ratio of the mass of the gas in the atmosphere to the total mass of the atmosphere. Figure \ref{fig:sub2a} shows the snapshots of temperature vs $CO_{2}$ and $CH_{4}$ used in our model and Figure \ref{fig:sub2b} shows this same data interpolated to create a 2D grid we can then use in our life-climate coupled model. We use this grid to look up the corresponding average surface temperature for any atmospheric composition throughout our experiments.

\subsection{Microbe setup} 
\label{Section:microbes}

We restrict life to a single species of single-celled methanogens that generate adenosine triphosphate (ATP), an organic compound that provides energy for cellular processes, via the following metabolism

\begin{equation}
\label{Eq:metabolism}
CO_{2}\ +\ 4H_{2} \rightarrow CH_{4}\ +\ 2H_{2}O,
\end{equation}

and creates biomass from

\begin{equation}
\label{Eq:biomass}
CO_{2}\ +\ 2 H_{2} (\ +\ ATP) \rightarrow CH_{2}O\ +\ H_{2}O.
\end{equation}

We explore two scenarios for the energy obtained from the microbes' metabolism, Equation \ref{Eq:metabolism}; the first, labeled energy scheme (a), where there is a fixed energy yield for metabolising 1 mole of $H_{2}$ representing a simplified case. This approach is typical in abstract models of life-environment coupled systems \citep[e.g.][]{williams2007flask, Nicholson:2018}. 

We then explore a second scenario, labeled energy scheme (b), where the energy yield from metabolising is determined by the free energy form of the Nernst equation following \cite{Kharecha:2005}, described by

\begin{equation}
\label{Equation:thermo}
\Delta G = \Delta G^{0} + R\ T\ log(Q)
\end{equation}

where $\Delta G^{0}$ free energy change of the reaction under standard conditions (i.e. unit concentrations of the reactants and products), $R$ = universal gas constant $= 0.008314 kJ\ mol^{-1}\ K^{-1}$, and $T = 298 K$ (the assumed water column temperature). $Q$ is calculated using

\begin{equation}
Q = \frac{\left[ CH_{4} \right]^{*}_{aq} \cdot a(H_{2}O)^{2}}{\left[ CO_{2} \right]^{*}_{aq} \cdot (\left[ H_{2} \right]^{*}_{aq})^{4}},
\end{equation}

where $[i]^{*}_{aq} = \frac{[i]_{aq}}{\alpha (i)}$ - the dissolved concentration of species $i$ divided by its Henry's law coefficient, and $a(H_{2}O)$ is assumed to be $1$ \citep{Kharecha:2005}. Following \cite{Kral:1998} we assume that $\Delta G^{0} = (-253 + 0.41\ T) kJ\ mol^{-1}$. We explore these two energetic scenarios to allow us to observe how the behaviour of our model changes with incremental increases to its complexity.

For the the experiments presented in this study there are no restrictions on the growth of methanogens other than the availability of $CO_{2}$ and $H_{2}$. We assume that microbes can uptake $H_{2}$ at a maximum rate of $C^{H_2}_{max}$ (see Table \ref{table4}), and no set minimum rate as long as there is $H_{2}$ in the ocean for the microbes to consume. Studies of microbes on Earth show that metabolic rates vary in response to the environment (e.g., competition for resources etc.) up to some maximum value based on studies of microbes in `ideal' laboratory conditions \citep{monod2012growth, brown2004toward,  li2019predictive}. In our model this fundamental principle is accounted for through the variation in $H_{2}$ uptake to a maximum rate. This approach follows the abstract representations of life used in abstract life-environment coupled models \cite[e.g.][]{williams2007flask, Nicholson:2018}. This means that in this model the microbes can draw the concentration of $H_{2}$ in the ocean down to zero. Real microbes are limited by diffusion of $H_{2}$ (and other required nutrients) across their cell membranes, and microbes have evolved a range of adaptations to maximise the rate of diffusion into their cell, from changing cell sizes and shapes, to growing filaments to maximise surface area, and developing strategies to move to areas of higher resource concentration \citep{beveridge:1988, siefert:1998, koch:1996, schulz:2001, young:2006}. Therefore real methanogens cannot draw $H_{2}$ concentrations to zero as they require a chemical gradient between themselves and their environment in order to uptake nutrients. We make the assumption in this study that the microbes have evolved to exploit $H_{2}$ to the same limit for every experiment presented, and for simplicity we take this limit to be zero. In this study we are interested in comparing different population dynamics of biospheres that otherwise interact with the abiotic environment in the same way, e.g. the same metabolic pathways and the same ability to exploit limiting nutrients. This allows us to isolate any dependence of the resulting biosignature on these population dynamics in a simple manner. For accurate predictions of biosigatures for real planets, understanding the limit to which a resource, in this case $H_{2}$, can be exploited by life will be important.

As we only explore scenarios where $H_{2}$ availability is the limiting factor on microbe growth as $CO_{2}$ is far more abundant than $H_{2}$ on the model planet, the uptake of $CO_{2}$ by the microbes is determined by the uptake of $H_{2}$. Therefore if microbes are consuming $H_{2}$ to generate biomass they will consume 1 mole of $CO_{2}$ for every 4 moles of $H_{2}$ following Equation \ref{Eq:metabolism}. If the microbes are instead generating ATP they will consume 1 mole of $CO_{2}$ for every 2 moles of $H_{2}$ following Equation \ref{Eq:biomass}. If instead we modeled scenarios where $CO_{2}$ was more scarce than $H_{2}$ then the uptake of $H_{2}$ would depend on the uptake of $CO_{2}$ instead.

Another simplification made in our model is that when the microbes die, their bodies are buried with 100$\%$ efficiency, i.e. no recycling takes place. This means that once a microbe dies, the $CH_{2}O$ that makes up its cell is removed from the system. This simplification allows us to avoid introducing additional chemical reactions for the breakdown of $CH_{2}O$ (although incorporating more complex ecosystems is something we plan to explore in future work). There are two ways microbes can die in our model - 1) starvation due to insufficient ATP for maintenance and 2) via a constant death rate that represents death from any other cause. This background death rate effectively sets an average lifespan for the microbes.

The model microbes impact their planet via their metabolism and biomass creation. They remove $H_{2}$ and $CO_{2}$ from the ocean and excrete $CH_{4}$. Due to the exchange of gases between the atmosphere and the ocean (Equation \ref{Equation:airseaexchange}) this then impacts the surface temperature of the planet (as detailed in Section \ref{Section:temp_CH4_CO2}).

Table \ref{table4} shows the default parameter values used for our model microbes. These values for the microbe cell maintenance, ATP cost for maintenance and cell growth, maximum uptake of $H_{2}$ per cell ($C^{H_{2}}_{max}$), and ATP generated via their metabolism, are used for our initial experiments and then these values are changed to measure the sensitivity of our model results on these parameters.

\begin{table*}
\caption{Model methanogen parameter values. $*$ from \protect\cite{Lynch:2019}, $\alpha$ from \protect\cite{Janssen:1996}.}
\begin{center}
\begin{tabular}{|c|c|} \hline
Parameter & Value \\ \hline \hline
Maintenance ATP cost per microbe & $2.16 \times 10^{-19} mol_{ATP}\ cell^{-1}s^{-1}$ $^{*}$  \\ \hline
Growth ATP cost to build one microbe & $4.23 \times 10^{-14} mol_{ATP}\ cell^{-1}$ $^{*}$  \\ \hline
Maximum rate of $H_{2}$ consumption per microbe - $C^{H_2}_{max}$ & $3.76 \times 10^{-17}\ mol_{H_{2}}cell^{-1}s^{-1}$ $^{*}$ \\ \hline
Microbe cell protein ($CH_{2}O$) content & $7.4\times10^{-15}\ mol_{CH_{2}O}\  cell^{-1}$ $^{*,\ \alpha}$  \\ \hline
Death rate of microbe population & $1\%$ of population per hour \\ \hline
Energy scheme (a) - moles of ATP obtained per mole of $H_{2}$ consumed &  0.15 $mol_{ATP}\ mol^{-1}_{H_{2}}$ $^{*}$ \\ \hline 
Energy scheme (b) -  energy required to form 1 mole of ATP &  $32.5 kJ\ mol^{-1}_{ATP}$ $^{*}$  \\ \hline 
\end{tabular}
\end{center}
\label{table4}
\end{table*}%

The protein ($CH_{2}O$) content of a cell is calculated assuming that $m_{cell} \approx 2\ m_{cell\ protein} = 4.44\times 10^{-13}g$ \citep{Janssen:1996, Lynch:2019}. As another simplification microbes starve instantly if they do not have sufficient ATP for maintenance within the current timestep \citep[although in reality it can take microbes far longer to die from starvation; dormancy is as a common strategy for microbes living under resource limitation,][]{Lennon:2011}.

A standard method to model populations of life is through agent based dynamics, where we simulate the actions of individual agents to understand the behaviour of a system, and this approach is often used in ecological models \citep{CHRISTENSEN:2002, GrimmRailsback:2013, Nicholson:2018}. However, as our populations grow large this approach is not computationally feasible. Therefore, as we are aiming to capture the bulk population properties we consider the population in terms of the total ATP contained within it. In some abstract models of life-environment coupled systems such as the Flask model \citep{williams2007flask} or the ExoGaia model \citep{Nicholson:2018}, microbes within the model accumulate biomass over the course of the experiment and they must `spend' a certain amount of biomass each timestep in an maintenance cost. Additionally, there are biomass thresholds for starvation and reproduction where microbes die when their biomass drops below the `starvation threshold', and reproduce when their biomass exceeds the `reproduction threshold'. For real microbes maintenance costs (the energy required by a microbe for processes other than biomass generation) are given in terms of ATP \citep{csengor:2013, Lynch:2019}. Therefore as we can consider microbe growth in terms of the biomass created per mole of ATP \citep{csengor:2013}; instead of tracking the biomass within our population we track the ATP contained within the population and assume a constant amount of biomass per microbe (see Table \ref{table4}). Therefore in our model a microbe will accumulate ATP, using some for maintenance as needed, until it accumulates enough ATP to generate sufficient biomass for a new microbe, at which point it reproduces. Therefore, to track the population dynamics of the microbes we require a calculation of the total ATP within the population, alongside that lost due to starvation and `spent' during maintenance and reproduction. Our treatment of these elements is described in the following sections.

\subsubsection{Distribution of ATP in population}

Biomass within microbe cells varies even within the same species \citep{cermak2017direct} and a normal distribution has been used to capture diversity in agent based models of microbes \citep{hellweger2009bunch}. We consider our microbe population in terms of ATP contained within the population and represent this as a normal distribution centered around $\mu_{s}$ - the mean ATP available to a cell, and with variance of $\sigma_{s} = 0.1  \mu_{s}$, an approach used in \cite{Nicholson:2017, nicholson2018alternative, Nicholson:2018}. Therefore for our microbe population, the number of microbes containing $x$ moles of ATP, $f_{s}(x)$, is given by

\begin{equation}
f_{s}(x) = \frac{p_{tot}}{\sigma_{s} \sqrt{2 \pi}} e^{-\frac{1}{2} \left( \frac{x - \mu_{s}}{\sigma_{s}}\right)^{2} }.
\end{equation}

where $p_{tot}$ is the total population of the biosphere at the start of the biological timestep. We can then use this distribution to calculate the number of microbes above or below certain thresholds, e.g. the number of microbes with sufficient ATP to reproduce. 

\subsubsection{Number to starve}

Each timestep microbes must `spend' a certain amount of ATP to maintain basic function, those with insufficient ATP starve to death. To determine the number of microbes that are below this threshold and thus die via starvation, $s_{s}(a_{m})$, we use the cumulative distribution function to determine the population with ATP levels lower than maintenance threshold $a_{m}$

\begin{equation}
s_{s}(a_{m}) = p_{tot} \int^{a_{m}}_{-\infty} f_{s}(x) \ \partial x\ = \frac{ p_{tot}}{2}\left[ 1 + erf \left( \frac{a_{m} - \mu}{\sigma \sqrt{2}} \right) \right].
\end{equation}

\subsubsection{Amount of ATP within starved population}

We also need to calculate the amount of ATP contained within microbes that are below the starvation threshold ($ a_{m}$). This quantity of ATP,  $h_{s}(a_m)$, is given by

\begin{equation}
h_{s}(a_m) =  p_{tot} \int^{a_m}_{-\infty} x f_{s}(x)\ \partial x\ =  p_{tot} \int^{a_m}_{-\infty} \frac{x}{\sigma \sqrt{2 \pi}} e^{-\frac{1}{2} \left( \frac{x - \mu}{\sigma}\right)^{2} }\ \partial x .
\end{equation}

To solve we, substitute

\begin{equation}
\lambda = \frac{x - \mu}{\sigma}, \ \ \ \ \ \frac{\partial \lambda}{\partial x} = \frac{1}{\sigma},\ \ \ \ c\ =\ \frac{a_m - \mu}{\sigma},
\end{equation}

to get

\begin{equation}\label{eq2}
h_{s}(a_m) =  \ - \ p_{tot} \frac{\sigma}{\sqrt{2\pi}} e^{-\frac{1}{2}c^2} \ + \  p_{tot} \ \frac{\mu}{2} \left( 1 + erf \left(  \frac{c}{\sqrt{2}} \right) \right).
\end{equation}

\subsubsection{ATP maintenance cost}

Once the microbes that die from starvation are removed from the population, the total ATP used for maintenance, $m_{s}$, used by the remaining population, $r_{s}$, assuming a maintenance cost per microbe of $a_{m}$, is given by

\begin{equation}
m_{s} = r_{s} \times a_{m}.
\end{equation}

\subsubsection{Number of microbes to reproduce}

Finally, microbes with sufficient ATP to reproduce, $a_{r}$, will do so asexually by splitting into two identical individuals. The number of microbes with sufficient ATP to reproduce, $r_{s}(a_{r})$, is calculated by

\begin{equation}
r_{s}(a_{r}) =  p_{tot} -  \ \frac{p_{tot}}{2}\left[ 1 + erf \left( \frac{a_{r} - \mu}{\sigma \sqrt{2}} \right) \right] .
\end{equation}

\section{Experiment setup}
\label{Section:experimentsetup}

In the following section we will present results for a number of experiments. In our model we step forwards in time solving the equations governing the system described previously. We employ two different timesteps, one for biological processes such as microbe death and reproduction, and another for abiotic processes such as $H_{2}$ input to the atmosphere. We set the biological timescale to be an hour, and the abiotic \text{red}{timestep} to be a year. Microbial growth is often measured in units of an hour \citep[e.g.][]{Weissman:2021} and so we choose this for the biological timescale. The climate in the UM simulation setup \cite{Eager:2022} used to determine the relationship between atmospheric composition and surface temperature for this model tends to stabilise over the order of years to tens of years (depending on the starting configuration), and so we choose a year for our abiotic timescale. in our model. For each experiment we first run the model without life until the ocean and atmospheric chemistry is in equilibrium which takes around $\approx 12,000$ years (here and throughout `years' refers to Earth years). Equilibrium is determined when the surface temperature of the planet stabilises to a constant. We then reset our time counter to zero and introduce life at $t = 20,000$ years. Once life is introduced we run the simulation for a further $40,000$ years after which we then force a total extinction of the biosphere to demonstrate how the planet surface chemistry reacts to the removal of life.

To compare results from different experiments, instead of artificially removing life during the experiment we instead let the biosphere persist (if the microbes can survive their environment) until the end of the experiment. We then average the last $5,000$ years of each experiment to investigate how changing various biological parameters, such as the microbes' death rate or ATP maintenance cost, impacts the resulting biosignature - the abundance of methane in the atmosphere.

Time in the model progresses in terms of hours and years in the following steps:

\begin{enumerate}
\item Update yearly:
\begin{enumerate}
\item Abiotic influx and outflux of $H_{2}$, $CO_{2}$ and $CH_{4}$
\item Planetary temperature determined from atmospheric chemical makeup
\end{enumerate}
\item Update hourly:
\begin{enumerate}
\item Exchange of gases between the atmosphere and ocean (Equation \ref{Equation:airseaexchange})
\item Calculate number of microbes that die
\item Calculate remaining microbes' ATP maintenance cost
\item Calculate number of microbes with sufficient ATP reproduce given sufficient $H_{2}$ and $CO_{2}$ availability
\item Microbes consume remaining $H_{2}$ and $CO_{2}$ uptake capacity to create ATP and excrete $CH_{4}$
\end{enumerate}
\end{enumerate}

When referring to changing parameters in some of the results presented in this paper, the default values for these parameters are those given in Table \ref{table1} and Table \ref{table4}. For each experiment a seed population size $p_{seed} = 10^{2}$ is used unless otherwise specified. However it is demonstrated in Section \ref{Section:equilibrium_state} that the model results are insensitive to the size of the seed population used.

In Section \ref{Section:H2_limited} we will explore scenarios where $H_{2}$ is the limiting resource on microbe growth. We use $H_{2}$ as our limiting resource as this mimics the scenarios for life on early Earth. Exploring scenarios for an $H_{2}$ dominated atmosphere would also involve including the greenhouse effect of abundant atmospheric $H_{2}$ \citep{Pierrehumbert:2011}. Instead we take $H_{2}$ as the limiting resource and calculate the average surface temperature based only on the atmospheric content of $CO_{2}$ and $CH_{4}$. This setup also means that the starting temperature of all our experiments is the same, even when changing the abundance of the limiting resource which makes for easier comparisons. 

The results presented in Section \ref{Section:H2_limited} show that life has a significant impact on the methane content of the atmosphere and thus the average surface temperature of its host planet. By excreting $CH_{4}$ the microbes raise the average surface temperature of their planet by roughly 5 degrees (K). We will demonstrate that for a simple $H_{2}$ limited biosphere significantly changing the microbe parameters, such as microbe death rate or maintenance ATP cost, does not significantly change the $CH_{4}$ level in the atmosphere due to the biosphere. Although the microbe population can change significantly when changing biological parameters, the resulting change in the total $CH_{4}$ output by the biosphere is much smaller. Changing the influx of $H_{2}$ to the system however has a much stronger impact on the level of $CH_{4}$ in the atmosphere as the increased availability of $H_{2}$ supports a higher population of microbes which in turns increases the total $CH_{4}$ output of the biosphere.

\section{Model results}
\label{Section:H2_limited}

Here we first present the results from a single experiment with parameters set to the default values found in Tables \ref{table1} and \ref{table4} in Section \ref{Section:microbes}. Initially the planet is devoid of life and at $t = 20,000$ years we seed the planet's ocean with our model microbial life. We then run the simulations with life for a further $40,000$ years, and then impose a total extinction at $60,000$ years, removing all life.

\begin{figure*}
\centering
\begin{subfigure}{.45\textwidth}
  \centering
  \includegraphics[scale=0.45]{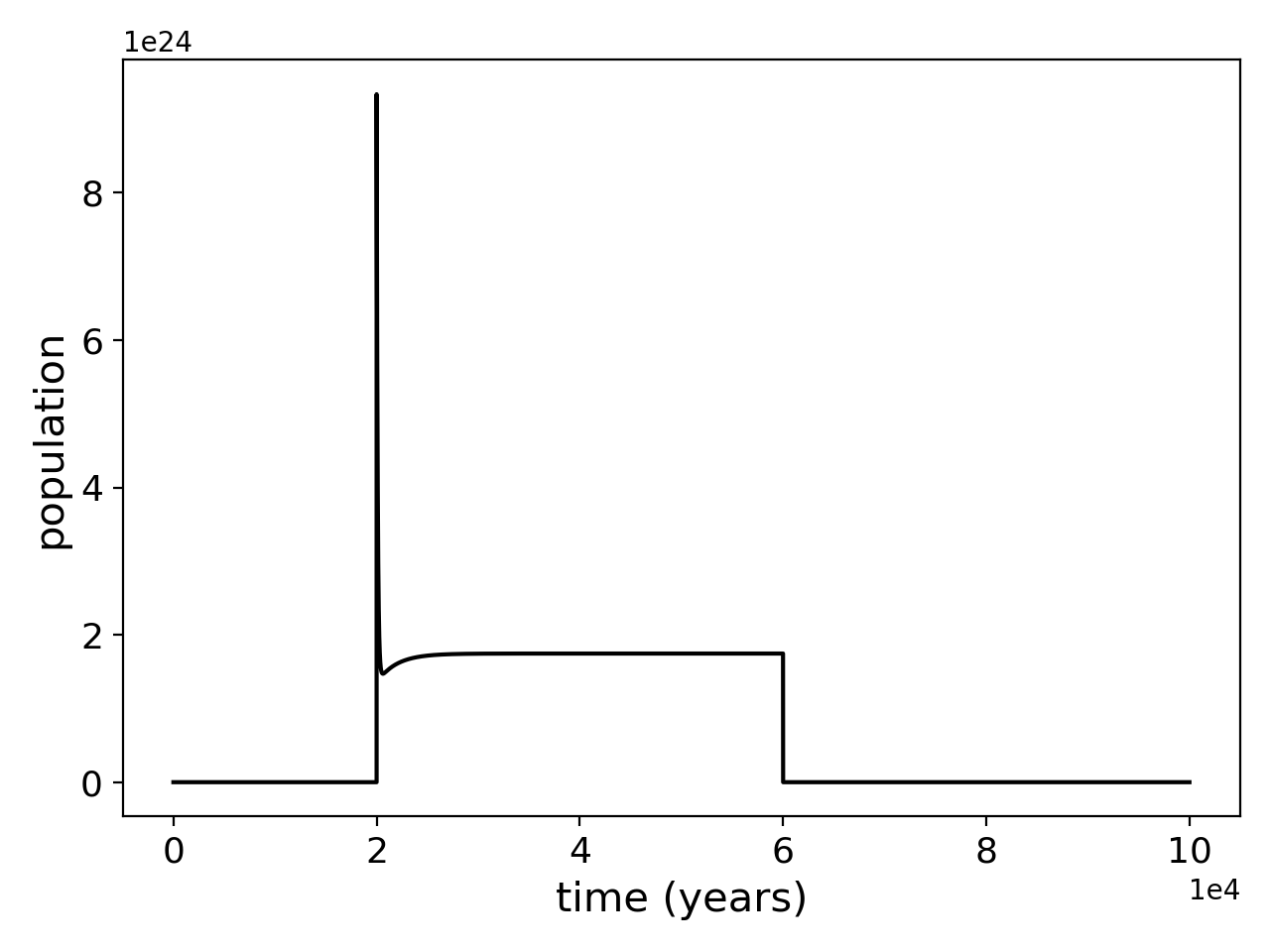}
  \caption{Microbe population}
  \label{fig5:b}
\end{subfigure}
\begin{subfigure}{.45\textwidth}
  \centering
  \includegraphics[scale=0.45]{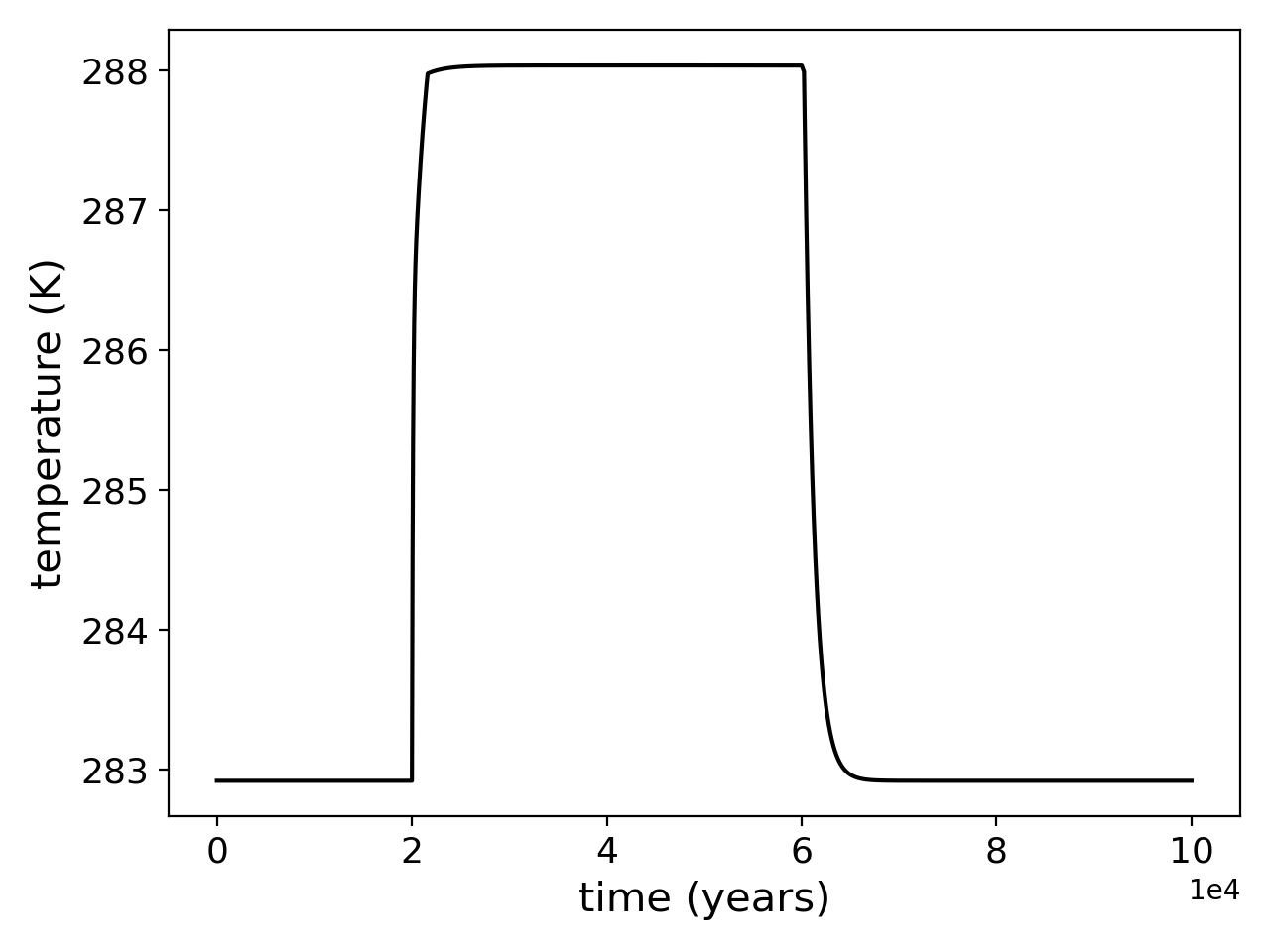}
  \caption{Average surface temperature (K)}
  \label{fig5:a}
\end{subfigure}%

\begin{subfigure}{.45\textwidth}
  \centering
  \includegraphics[scale=0.45]{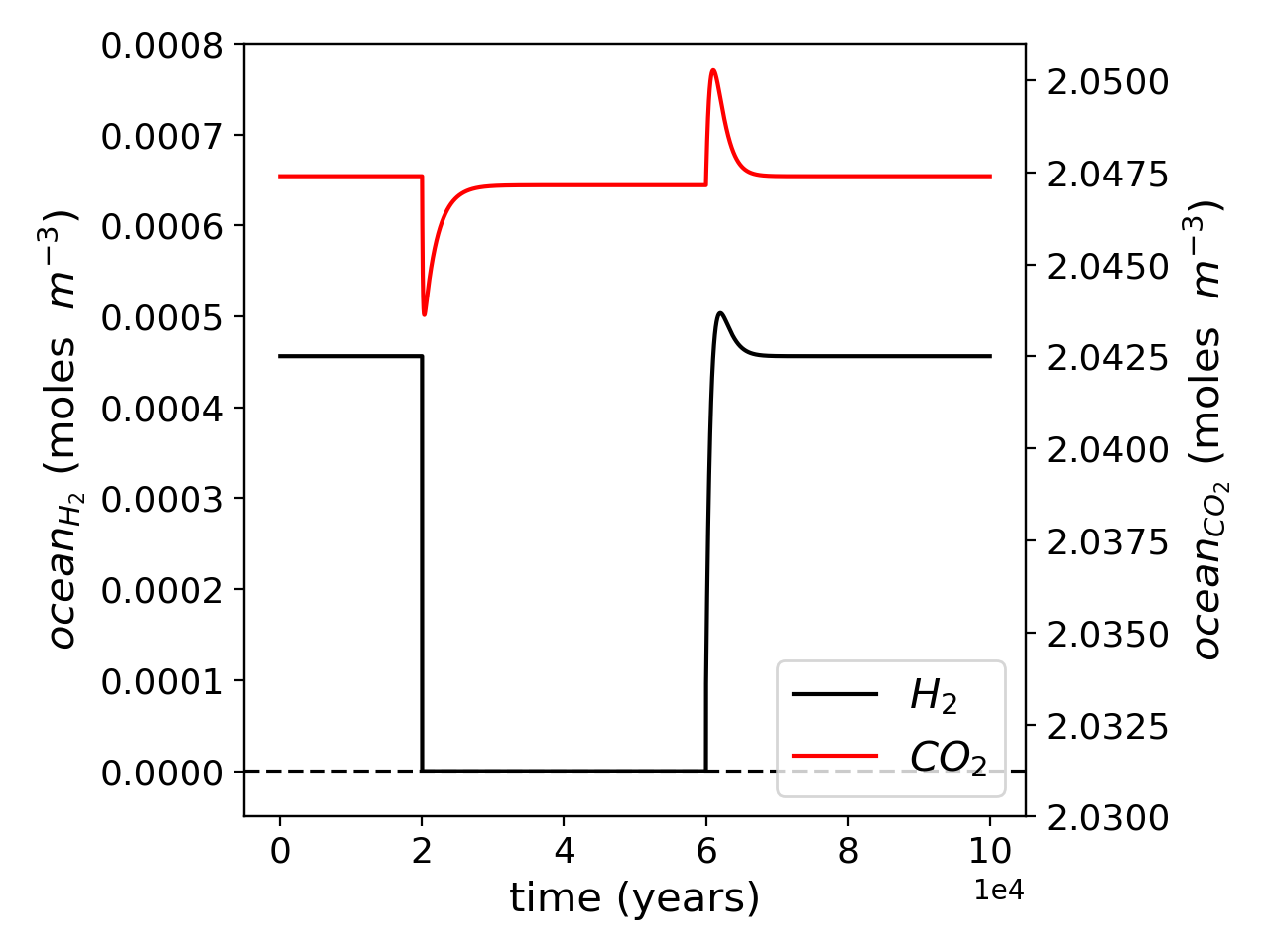}
  \caption{Ocean concentrations of $CO_{2}$ and $H_{2}$}
  \label{fig5:e}
\end{subfigure}%
\centering
\begin{subfigure}{.45\textwidth}
  \centering
  \includegraphics[scale=0.45]{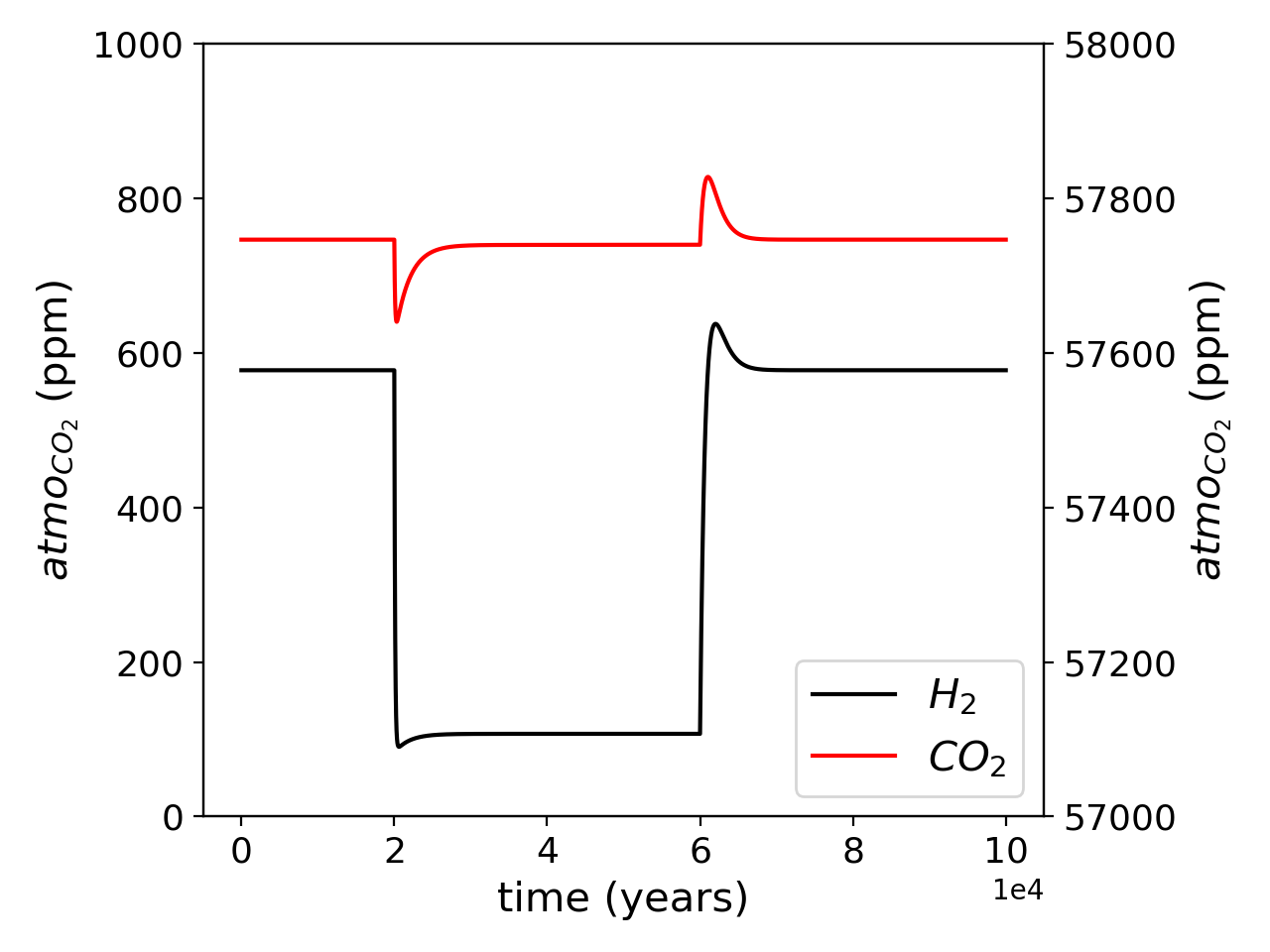}
  \caption{Atmospheric levels of $CO_{2}$ and $H_{2}$}
  \label{fig5:c}
\end{subfigure}%

\begin{subfigure}{.45\textwidth}
  \centering
  \includegraphics[scale=0.45]{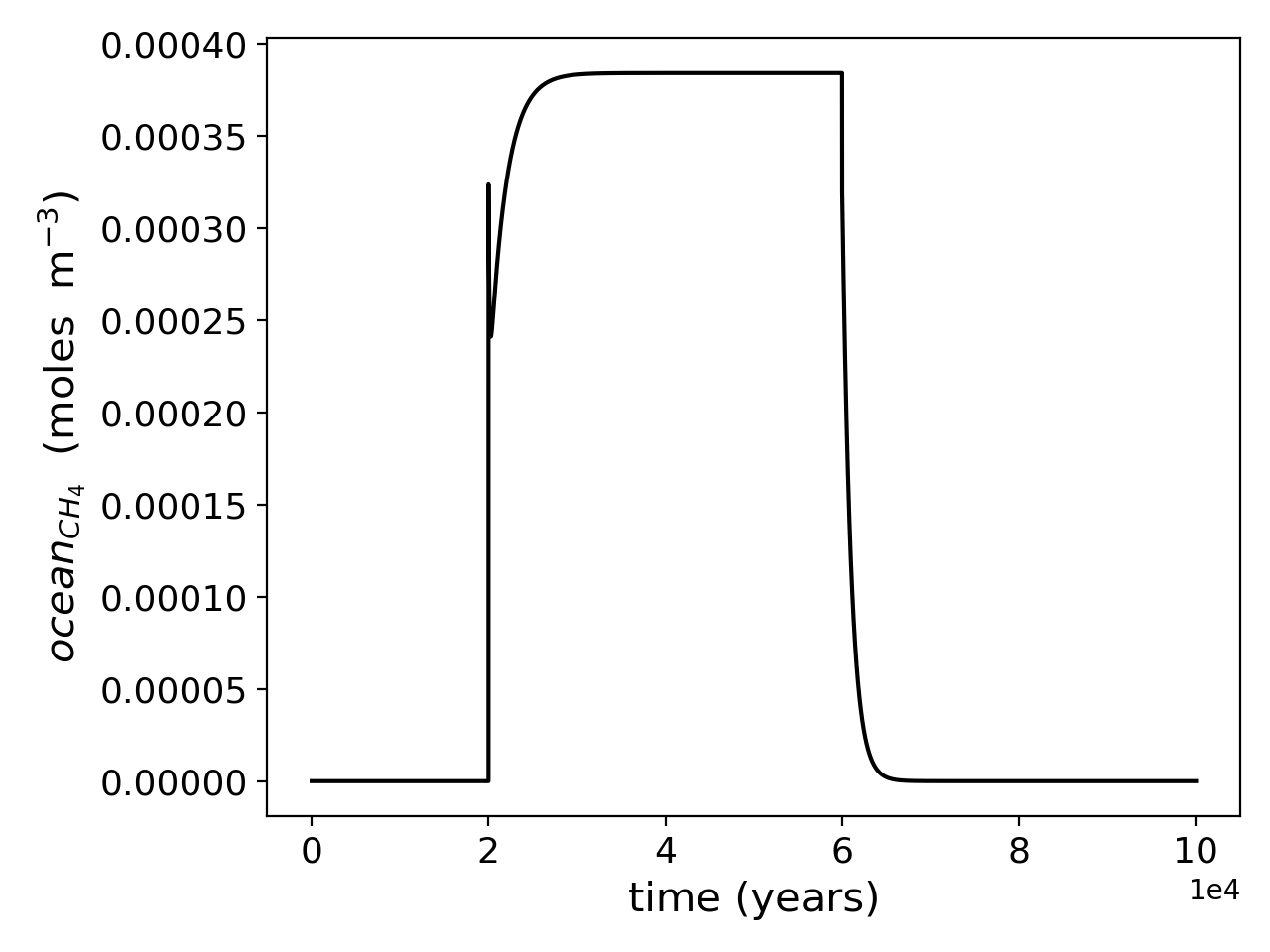}
  \caption{Ocean concentration of $CH_{4}$}
  \label{fig5:f}
\end{subfigure}
\begin{subfigure}{.45\textwidth}
  \centering
  \includegraphics[scale=0.45]{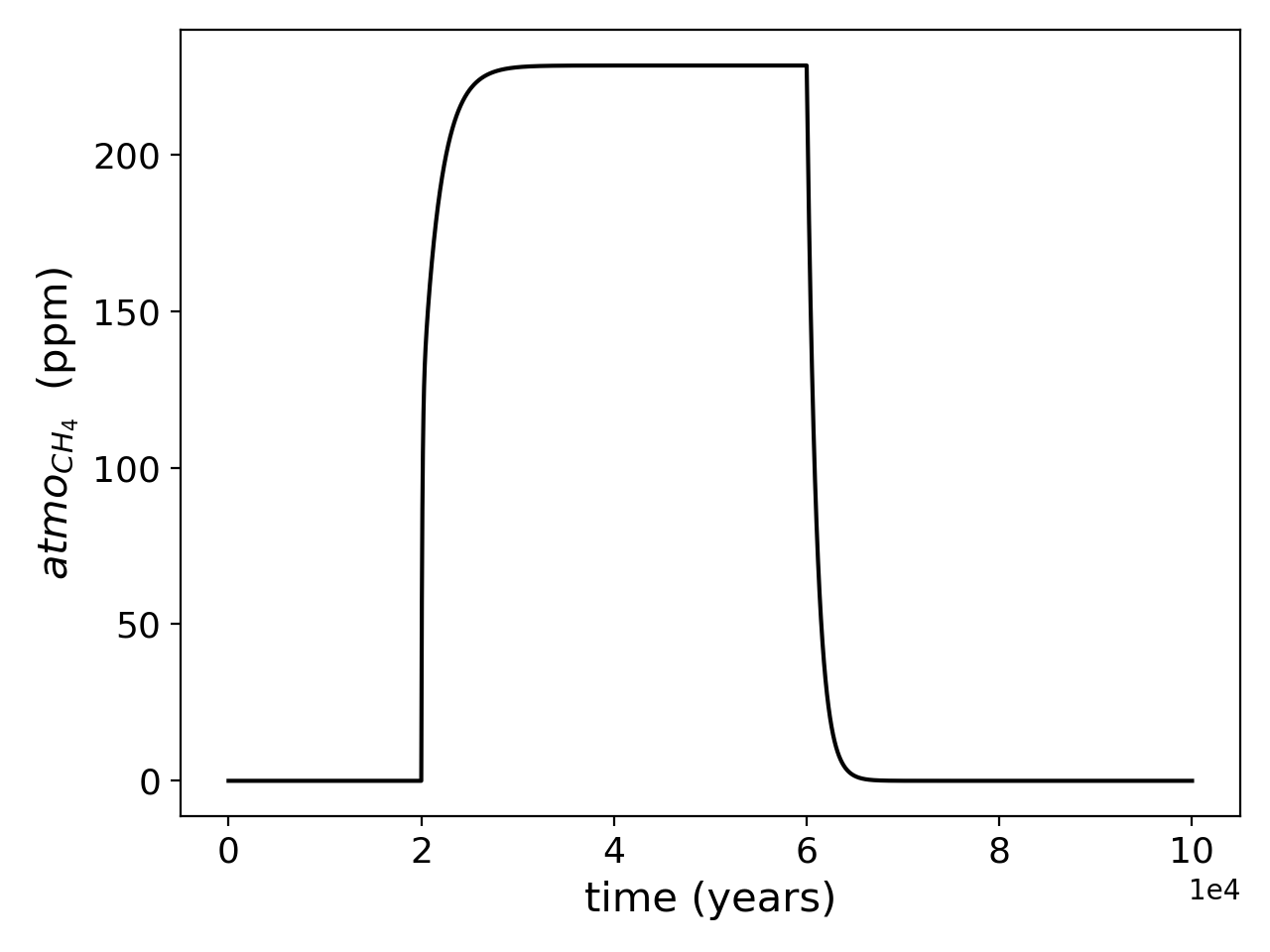}
  \caption{Atmospheric level of $CH_{4}$}
  \label{fig5:d}
\end{subfigure}
\centering
\caption{Model results for a single experiment showing how the planet's surface temperature, and the ocean and atmospheric levels of $H_{2}$, $CO_{2}$ and $CH_{4}$ change with the introduction of life at $t = 20,000$. The dashed line in panel c) shows a concentration of zero of $H_{2}$ dissolved in the ocean.}
\label{fig2:test}
\end{figure*}

Figure \ref{fig2:test} shows the output of a single run of our model. Figure \ref{fig5:b} shows the microbe population over time, \ref{fig5:a} shows the average surface temperature of the planet. Figures \ref{fig5:e} and \ref{fig5:f} show the abundance of $CO_{2}$, $H_{2}$ and $CH_{4}$ and in the ocean and Figures \ref{fig5:c} and \ref{fig5:d} show these abundances for the atmosphere. The impact of the appearance of life on the planet is immediately apparent in all six panels of Figure \ref{fig2:test}. Figure \ref{fig2:test} also shows that when life is removed from the planet that the atmospheric and ocean chemistry rapidly, in geological timescales, return to their previous states after roughly 5,000 years. Therefore in this model any biosignature produced by life is short lived if life goes extinct.

Figure \ref{fig5:b} shows that the microbe population experiences an initial spike before dropping to a stable population soon after life emerges. This initial spike in the total population is due to microbes emerging on a planet with abundant hydrogen dissolved in the ocean built up over time. Initially microbes are able to consume hydrogen at their maximum rate $C_{H_{2}}^{max}$ and this leads to exponential growth and a population explosion. As the population grows the biosphere draws down the level of $H_{2}$ in the ocean until they exhaust the available $H_{2}$. At this point the total population the planet can support is constrained by the influx of $H_{2}$ to the system (in this model provided by a representation of volcanic activity). As this is less than the concentration of $H_{2}$ that had built up in the ocean before life was introduced, the population rapidly declines until a stable state is reached where the reproduction rate of the microbes matches the death rate. The establishment and maintenance of this stable state once life is introduced to the planet is explored further in Section \ref{Section:equilibrium_state}.

Figures \ref{fig5:e} and \ref{fig5:f} shows a rapid transition in the ocean abundances of $H_{2}$, $CO_{2}$ and $CH_{4}$ after life appears on the planet. The ocean concentrations of $H_{2}$ and $CO_{2}$ are rapidly drawn down as life consumes these chemicals to generate biomass and ATP and $CH_{4}$ rapidly accumulates in the ocean as microbes excrete this gas as a byproduct of their metabolism (Equation \ref{Eq:metabolism}). We see that after the initial reduction in the ocean concentration of $CO_{2}$, $CO_{2}$ in the ocean rises again and stabilises to a level slightly below the initial concentration. This rise in the concentration of $CO_{2}$ in the ocean shortly after the drop corresponding to introduction of life is due to methane breakdown taking place in the atmosphere. Methane is recycled back to $CO_{2}$ and $H_{2}$ via Equation \ref{Equation:methanbreakdown} and so this in effect adds an extra input of $CO_{2}$ to the atmosphere on top of the influx from volcanic activity. We see in Figure \ref{fig5:e} that after the introduction of life, the $H_{2}$ concentration in the ocean remains at zero while life persists on the model planet. The asymmetry between the behaviour of ocean levels of $CO_{2}$ and $H_{2}$ is due to the differences between the removal of $H_{2}$ and $CO_{2}$ from the atmosphere, see Table \ref{table1}. Whereas the rate of $CO_{2}$ removal in our model depends only on the abundance of $CO_{2}$ in the atmosphere, $H_{2}$ loss depends on both the abundance of $H_{2}$ and $CH_{4}$ in an approximation of the process of $H_{2}$ loss to space on real planets. As the level of $H_{2}$ in the atmosphere decreases, the level of $CH_{4}$ rises, and so the loss of $H_{2}$ happens at a faster rate than if it were just dependant on atmospheric $H_{2}$. The availability of $H_{2}$ sets the total population that the planet can support and so in our model the microbes reproduce until $H_{2}$ is depleted in the ocean down to a concentration of zero. If $CO_{2}$ was instead the limiting resource in our model, the ocean concentration of $CO_{2}$ would instead be drawn down to zero.

The behaviour of the abundance of $CH_{4}$, $CO_{2}$ and $H_{2}$ in the atmosphere largely follows what we see in the ocean. After life emerges $H_{2}$ and $CO_{2}$ are drawn down, levels of $CH_{4}$ rise, and we see a rise in atmospheric $CO_{2}$, after the initial drop, due to the recycling of methane. A much smaller rise is seen in the level of atmospheric $H_{2}$ because of the different process of $H_{2}$ removal.  We see in Figure \ref{fig5:a} that the emergence of life leads to a rapid 5 degree increase in the temperature of the planet due to the methane building up in the atmosphere as a byproduct of the microbes' metabolism. 

\subsection{Equilibrium state}
\label{Section:equilibrium_state}

The maximum growth rate of the model microbes is set by $C_{max}^{H_{2}}$ - the maximum rate at which the microbes can uptake $H_{2}$. The value for $C_{max}^{H_{2}}$ in Table \ref{table4} is based on studies of microbes in `ideal' laboratory conditions, in the absence of resource limitation or competition \citep{monod2012growth, wang2016mechanistic, Lynch:2019, li2019predictive}. $C_{max}^{H_{2}}$ represents the maximum metabolic rate for a microbe. However in realistic scenarios the metabolic rate of microbes changes due to environmental conditions such as resource shortages \citep{brown2004toward}. In our model, $C_{max}^{H_{2}}$ sets the maximum metabolic rate of the microbes however their metabolic rate can vary depending on nutrient availability. For a high metabolic rate where the birth rate of microbes exceeds the death rate, population growth will occur, and if their metabolic rate drops such that the growth rate is less than the death rate the microbe population will shrink and if the low growth rate persists total extinction will occur. A stable population is achieved when the growth rate of the microbes is equal to the death rate.

The population at which the biosphere stabilises at is determined by a feedback loop between $H_{2}$ availability per microbe, and microbe growth rate. Initially when microbes are seeded onto a model planet initially there will be abundant $H_{2}$ and microbes will uptake $H_{2}$ at their maximum rate - $C^{H_{2}}_{max}$ (see Table \ref{table4}). Microbes will be able to quickly accumulate ATP via Equation \ref{Eq:metabolism} and exponential growth of microbes will occur. This causes the large spike in microbe population soon after seeding seen in Figure \ref{fig5:b}. As the microbe population grows, the $H_{2}$ concentration in the ocean will fall (Figure \ref{fig5:e}).

\begin{figure}
    \centering
    \includegraphics[scale=0.35]{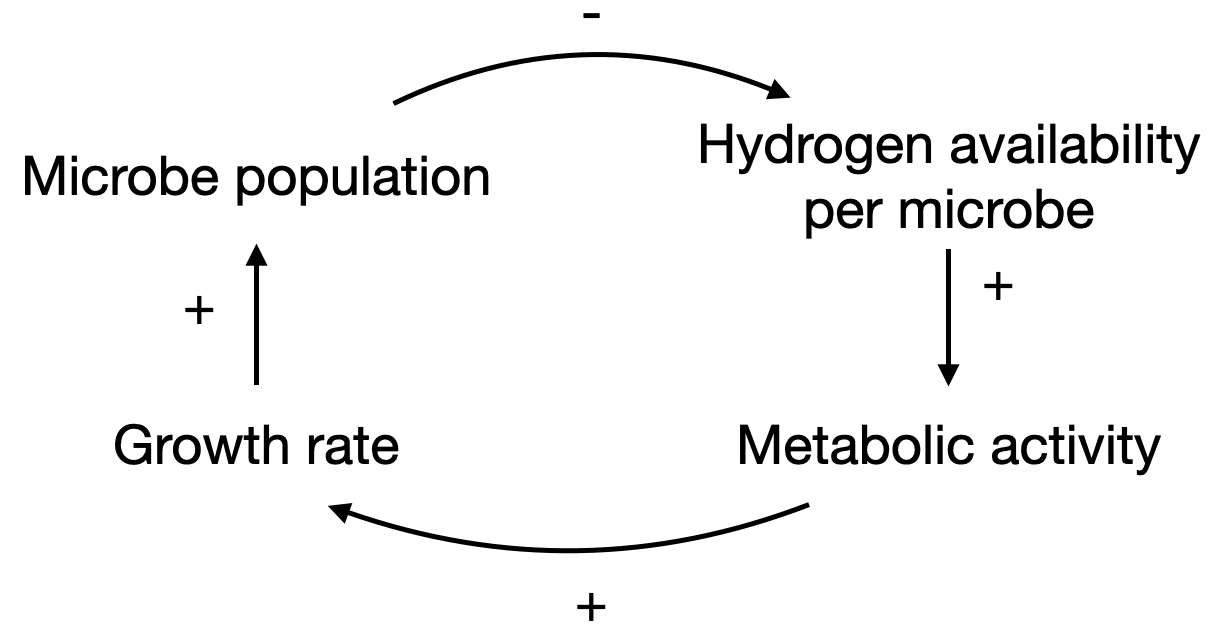}
    \caption{Diagram showing the feedback loop between $H_{2}$ availability per microbe, microbe metabolic rate, microbe growth rate and total microbe population. A + sign indicates that an increase in the source leads to an increase in the sink. A - sign indicates that an increase in the source leads to a decrease in the sink.}
    \label{fig:feedback}
\end{figure}

A point will be reached where the initial rapid growth of the microbes can no longer be supported by the diminishing $H_{2}$ content of the ocean, and the growth of microbes will become limited by the inflow of $H_{2}$ to the ocean from the atmosphere. This $H_{2}$ inflow rate is significantly smaller than the reserves of $H_{2}$ that had dissolved into the ocean before the emergence of life and so the high microbe population resulting from the initial exponential growth cannot be supported by the model planet long term. This causes the microbe population to drop. The microbe population will stabilise where the growth rate of the microbes is equal to the death rate of the microbes (Table \ref{table4}). The growth rate of a microbe depends on the availability of $H_{2}$, the ATP maintenance cost of the microbe and the cost of biomass production. If the availability of hydrogen increases, then the metabolic activity of the microbes can increase leading to an increase in the microbes' growth rate and so an increase in the total microbe population. An increase in the total population however will decrease the availability of hydrogen per microbe and so the feedback loop is stabilising. This feedback loop is shown in Figure \ref{fig:feedback}.

The microbe population will stabilise at a level where, on average, the $H_{2}$ availability per microbe is sufficient for that microbe to reproduce once before its death. In this way a stable population emerges in our model.

\begin{figure}
    \centering
    \includegraphics[scale=0.49]{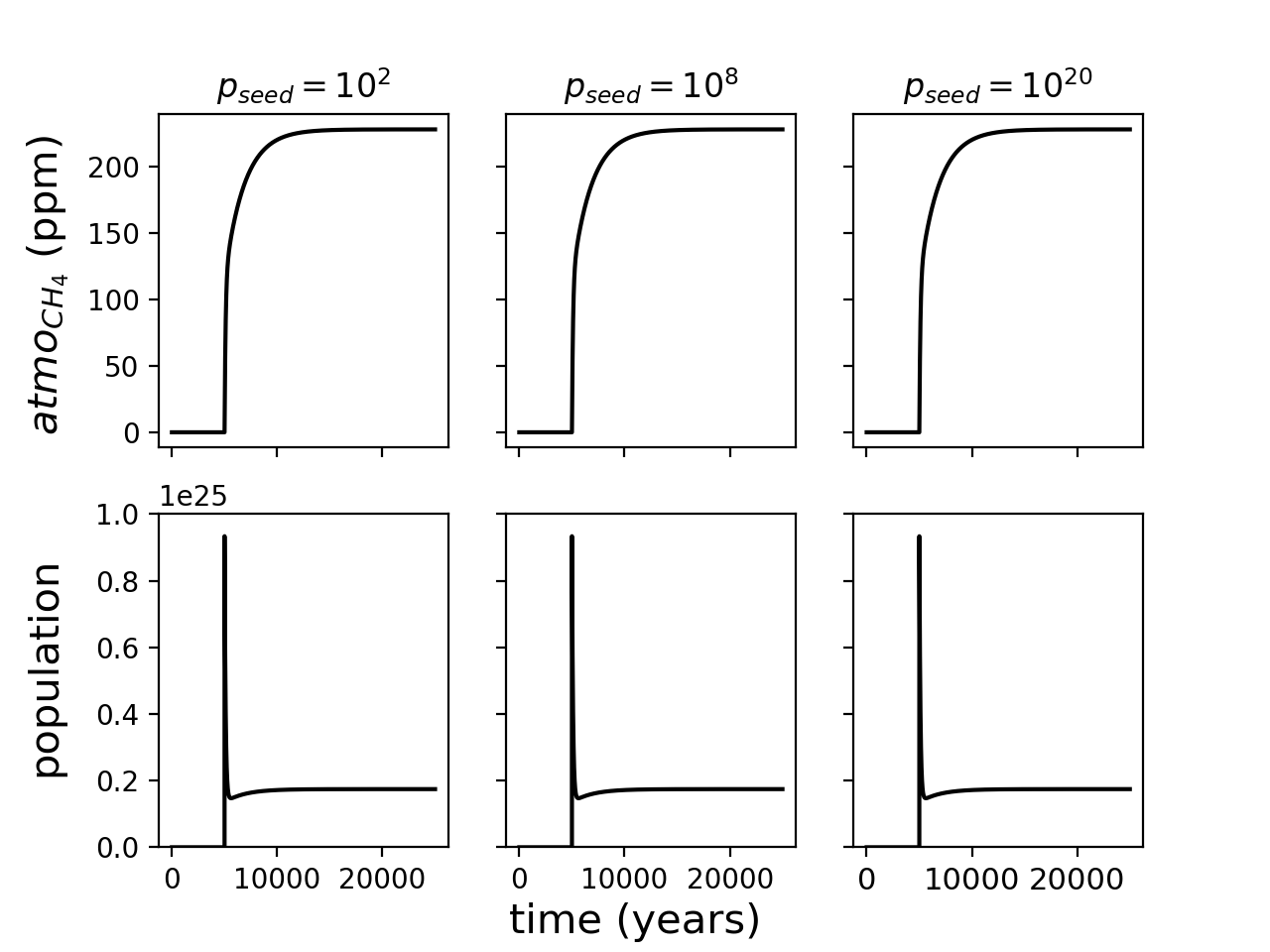}
    \caption{Figures showing the insensitivity of the total population, and the level of $CH_{4}$ in the atmosphere ($atmo_{CH_{4}}$), to the size of the microbial seed population $p_{seed}$.}
    \label{fig:seed_pop}
\end{figure}

The model results are insensitive to the size of the seed population. The initial spike in total population seen upon life emerging on a planet will vary minimally depending on the seed population, but as the stable population is determined by the inflow of $H_{2}$ to the ocean, which itself is determined by the influx of $H_{2}$ to the atmosphere, the stable population size reached is unaffected by the seed population size, see Figure \ref{fig:seed_pop}.

\subsection{Parameter Sensitivity}
\label{Section:parametersensitivty}

The parameters used in the model so far are taken from the values found in Table \ref{table4} which are based on measurements of methanogens grown in lab settings. However methanogens on Earth are highly diverse with different variants having different growth rates \citep{lyu2018methanogenesis}. Therefore, the parameters in Table \ref{table4} could clearly be very different for alien life on another planet as they differ significantly between microbes species on Earth. To investigate the sensitivity of our model results to different microbe parameters we changed the death rate (the percentage of microbes removed from the ocean per hour), microbe cell protein content, and the ATP maintenance cost per microbe, and repeated the experiment. These experiments reveal that the impact of microbes on the wider planetary system is only weakly dependent on these underlying characteristics of the microbes, while a viable biosphere is possible on the planet under the biological parameters chosen for the experiment. 

\subsubsection{Changing the death rate}
\label{Section:death-rate}

We define the death rate as the percentage of microbes that are removed per hour. In our model as there is no breakdown of $CH_{2}O$ - the building block for our life - microbe bodies are assumed to be removed from the system once dead. We find that while significantly changing the death rate leads to a large change in the total microbe population, it leads to only a small change in the abundance of methane in the atmosphere.

\begin{figure}
    \centering
    \begin{subfigure}{0.49\textwidth}
    \includegraphics[scale=0.53]{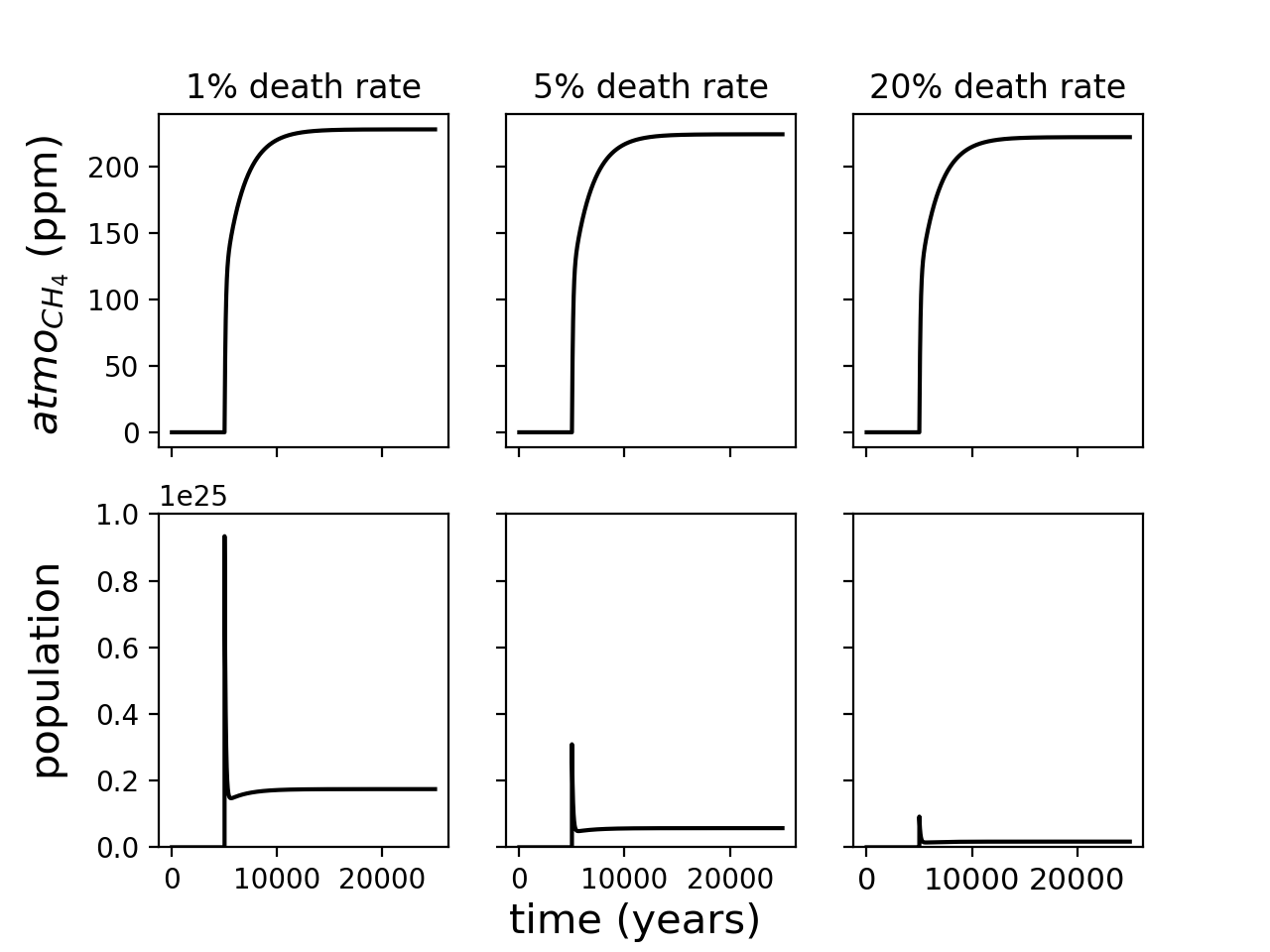}
    \caption{Impact of changing death rate on the atmospheric methane content ($atmo_{CH_{4}}$) and microbe population.}
    \label{fig:my_label11a}
    \end{subfigure}
    \begin{subfigure}{0.48\textwidth}
    \includegraphics[scale=0.53]{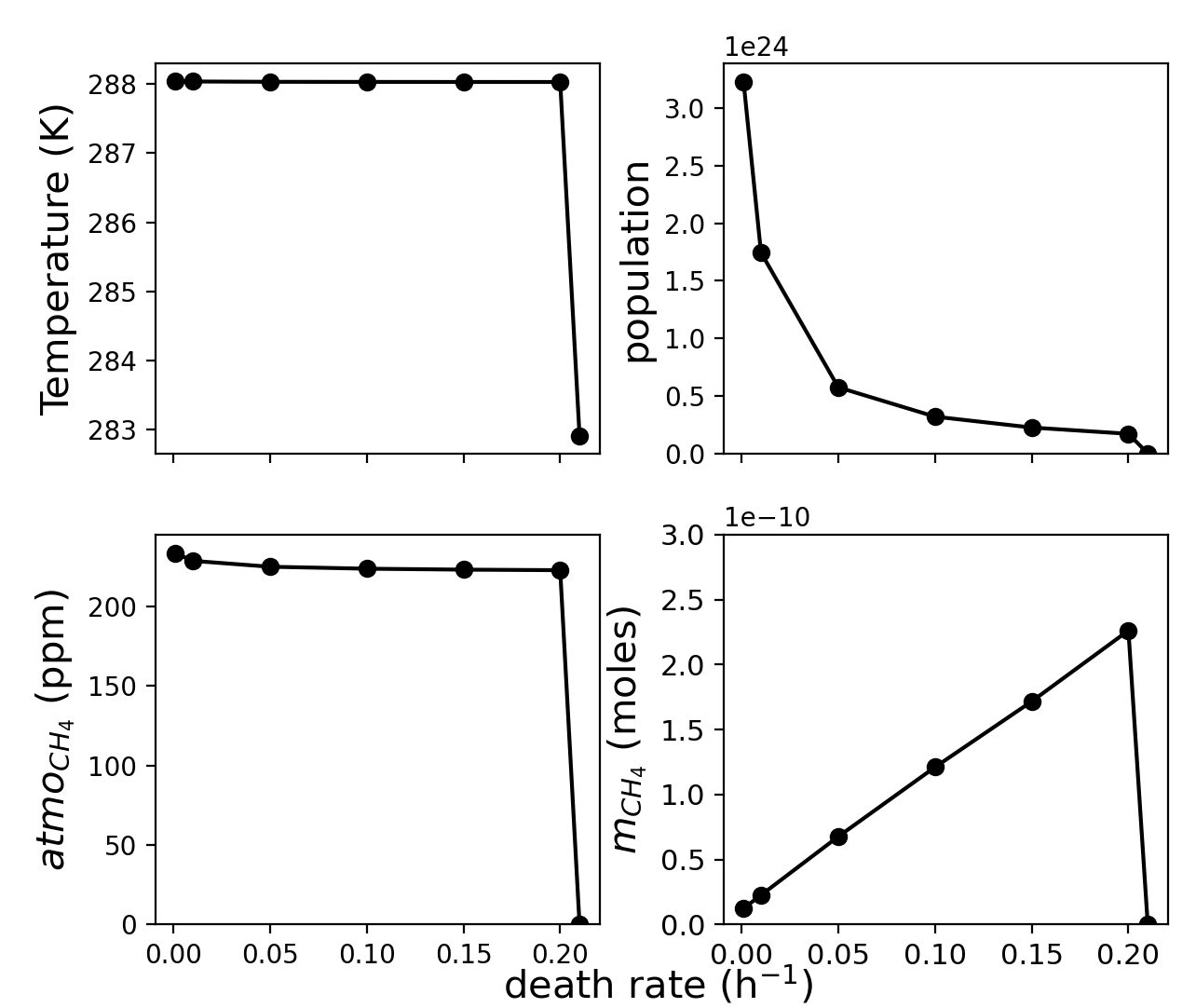}
    \caption{Sensitivity of the surface temperature, microbe population, atmospheric $CH_{4}$ content ($atmo_{CH_{4}}$), and the average $CH_{4}$ output per microbe ($m_{CH_{4}}$) to changing microbe death rate.}
    \label{fig:my_label11b}
    \end{subfigure}
    \caption{Panels showing the model sensitivity to changing the death rate of the microbes (determined as the fraction of the population removed per hour).}
    \label{fig:my_label11}
\end{figure}

Figure \ref{fig:my_label11} shows in panel (a) the level of methane in the atmosphere and microbe population over time for differing death rates, and in panel (b) the surface temperature, microbe population, atmospheric $CH_{4}$ content, labeled $atmo_{CH_{4}}$, and the average $CH_{4}$ output per microbe per year, labeled as $m_{CH_{4}}$, for varying death rates. We explored a range of death rates from $0.1\%$ up to $15\%$ of the population removed every hour. In Figure \ref{fig:my_label11b} we show the death rate as the fraction of the population removed per hour instead of the percentage. We see in Figure \ref{fig:my_label11} that a death rate of over 20$\%$ leads to extinction on the planet and no biosignature. 

Figure \ref{fig:my_label11a} demonstrates how the scale of the impact on the microbe population, and the impact on the abundance of atmospheric methane, differ significantly when adjusting the death rate. Figure \ref{fig:my_label11a} shows a clear impact on the microbe population as we change the death rate, as the death rate increases the population decreases. As more microbes are being removed regularly at a higher death rates this an unsurprising result. The impact on the methane in the atmosphere however is far more subtle. A large change in the population leads to only a small change in the abundance of atmospheric methane. Therefore changing the population dynamics significantly doesn't lead to a significant impact on the resulting biosignature. 

Figure \ref{fig:my_label11b} demonstrates the impact on the level of methane in the atmosphere from changing the death rate more clearly and shows that increasing the death rate leads to a small non-linear decrease in atmospheric methane. Due to the non linear relationship between methane in the atmosphere and surface temperature, we find that the temperature of the planet doesn't change significantly between each experiment. When we examine the average amount of methane output per microbe per hour - $m_{CH_{4}}$ - we see why a large change in the population translates to only a small change in the level of atmospheric methane, as we increase the death rate, $m_{CH_{4}}$ increases. For a higher death rate, the metabolic activity of the average microbe must occur at a faster rate to maintain a stable population. If microbes are being removed from the ocean at a higher rate then the reproduction rate of the microbes must increase to counteract it; this requires microbes to generate ATP faster and therefore increases the $CH_{4}$ output per microbe, as methane is a byproduct of the microbes' metabolisms. Therefore a lower population results in a similar atmospheric methane level as each individual microbe is now outputting $CH_{4}$ at a faster rate. 

As the maximum rate of $H_{2}$ uptake, $C_{max}^{H_{2}}$, is kept constant for all experiments, a point is reached where, with increasing death rate, it is no longer possible for the microbes to accumulate sufficient ATP to reproduce before they are killed off. Past this point a stable population isn't possible and the biosphere rapidly goes extinct.

\subsubsection{Changing additional biological parameters}
\label{Section:additionalbiologicalparameters}

We also investigated changing the ATP maintenance cost of the microbes, and their cell protein content (number of moles of $CH_{2}O$ per microbe) and found that, similar to changing the death rate, these parameters had only a small impact on the abundance of methane in the atmosphere despite having large impacts on the total population, up until the point where the biosphere collapses due to an insufficient $C_{max}^{H_{2}}$ for microbe growth. 

\begin{figure}
    \centering
    \includegraphics[scale=0.53]{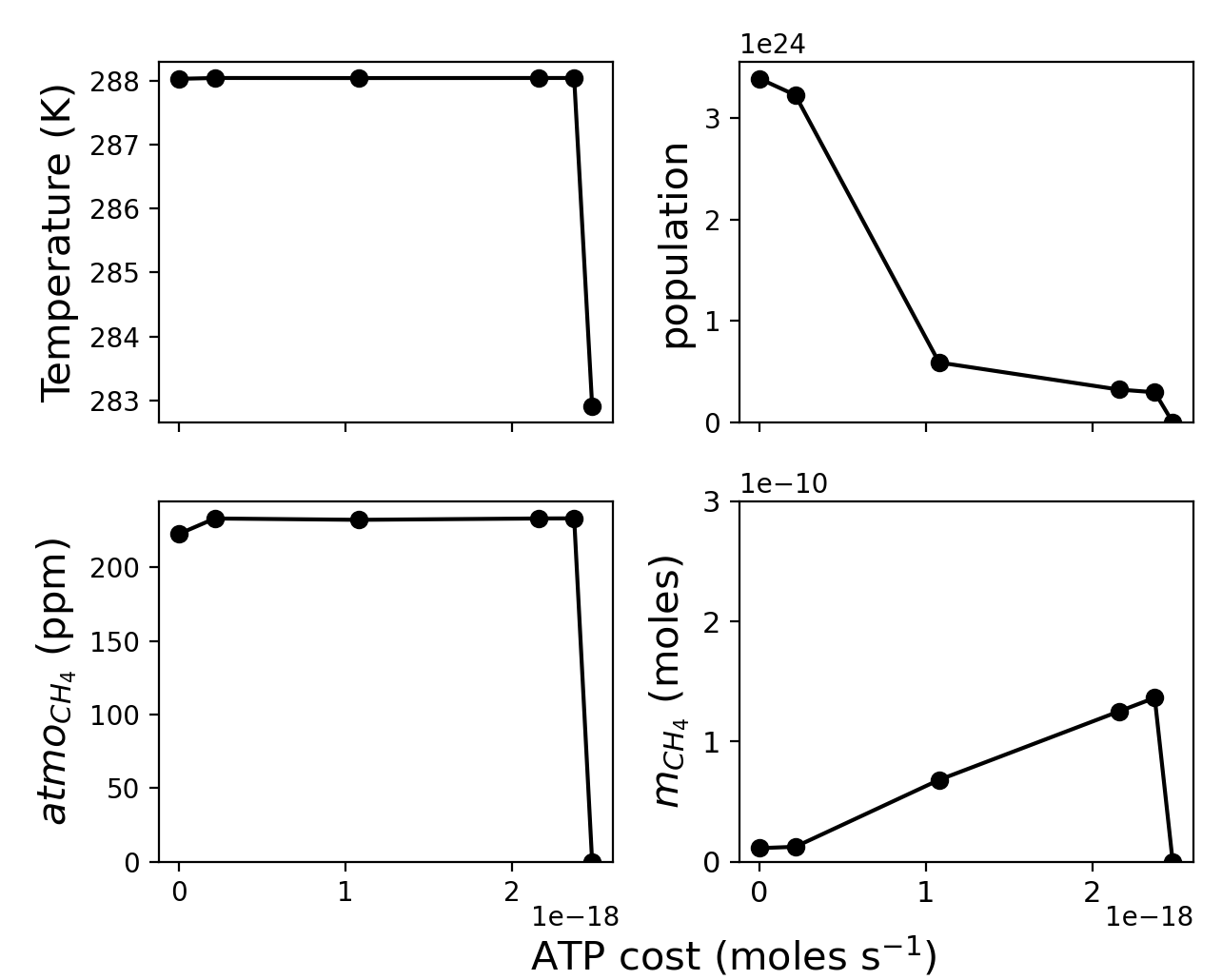}
    \caption{
    Panels showing the sensitivity of the surface temperature, microbe population, atmospheric $CH_{4}$ content ($atmo_{CH_{4}}$), and the average $CH_{4}$ output per microbe ($m_{CH_{4}}$) to changing the microbe ATP maintenance cost.}
    \label{fig:my_label13}
\end{figure}

Figure \ref{fig:my_label13} shows how the surface temperature, microbe population, level of methane in the atmosphere, and the average methane output per microbe $m_{CH_{4}}$, change with varying the ATP maintenance cost of the microbes. This is the ATP microbes must `spend' per second (measured in seconds as this is typical for laboratory measurements of microbes) to maintain basic functions and avoid death via `starvation' (this is separate from the imposed death rate explored in Section \ref{Section:death-rate}). We explored a range of maintenance costs from 0.01 times the value in Table \ref{table4} up to 10 times this value. In our experiments any microbe with insufficient ATP for this maintenance cost dies immediately and is removed from the system. All other microbe parameters in these experiments, including the death rate, were set to the values found in Table \ref{table4}. 

We find that increasing the ATP cost corresponds with no measurable change to the surface temperature of the planet, a slight increase in the atmospheric level of $CH_{4}$, a significant decrease in the microbe population, and again an increase in the value of $m_{CH_{4}}$ - the average methane output per microbe per year. Increasing the ATP cost per microbe requires the microbes' metabolic rate to increase in order to generate sufficient ATP to maintain a stable population. With microbes `spending' more of their $H_{2}$ and $CO_{2}$ intake on ATP production, there is less available for biomass production and so the total population the planet can support is reduced. The increased metabolic activity means that despite the reduced microbe population the amount of methane in the atmosphere rises only slightly with increasing ATP cost.

\begin{figure}
    \centering
    \includegraphics[scale=0.53]{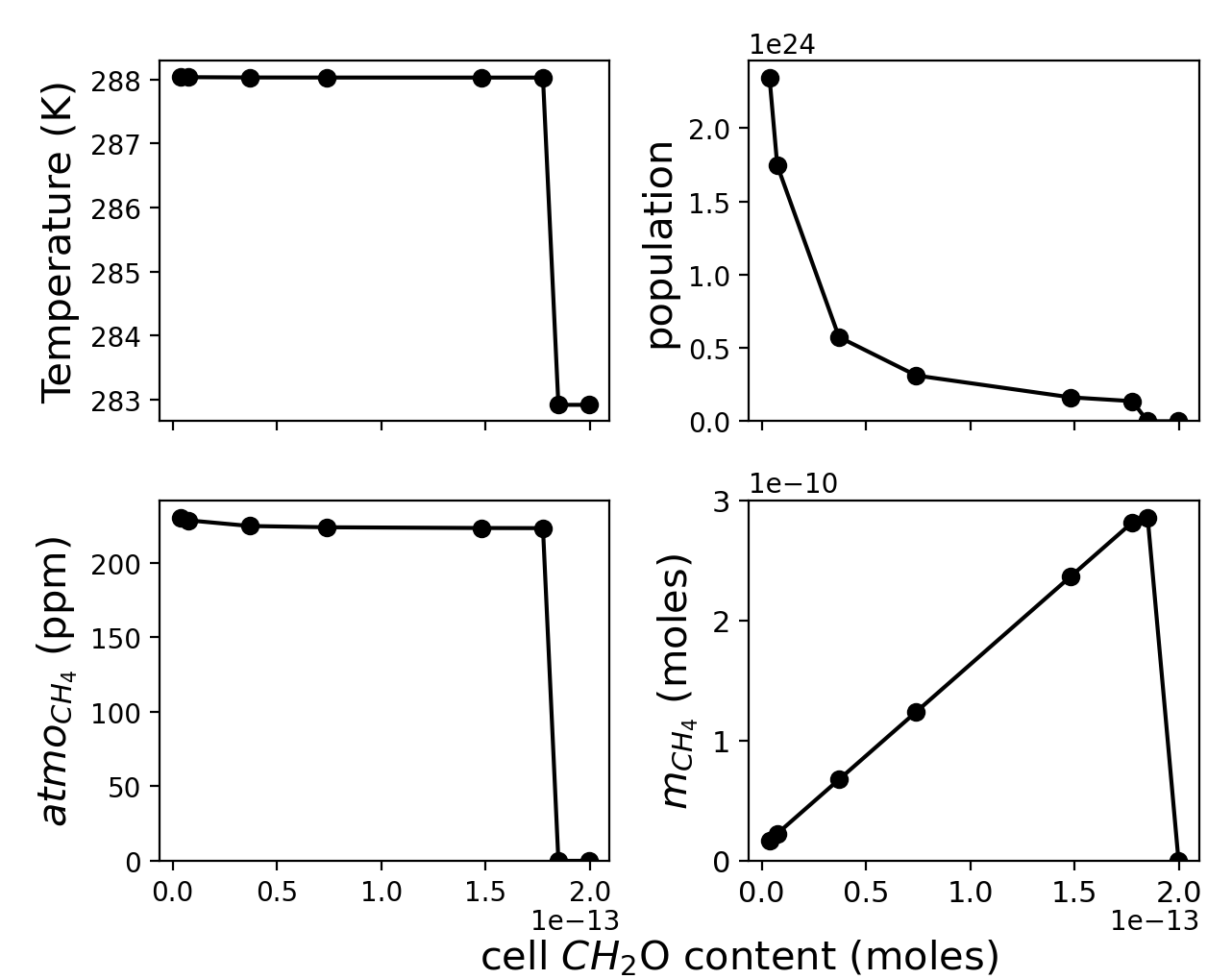}
    \caption{Panels showing the sensitivity of the surface temperature, microbe population, atmospheric $CH_{4}$ content ($atmo_{CH_{4}}$), and the average $CH_{4}$ output per microbe ($m_{CH_{4}}$) to changing the $CH_{2}O$ microbe cell content. Note that the ATP required to create a new cell scales linearly with cell $CH_{2}O$ content.}
    \label{fig:my_label14}
\end{figure}

Figure \ref{fig:my_label14} shows how the surface temperature, the microbe population, the atmospheric level of $CH_{4}$, and the average methane output per microbe per year ($m_{CH_{4}}$) change with varying the protein ($CH_{2}O$) content of the microbe's cells. As the cell size increases we increased the ATP `cost' to form a cell linearly. We varied the $CH_{2}O$ content of the cell from half of the value in Table \ref{table4} to over 25 times this value. We find again a point at which the cell size becomes too large and total extinction occurs as microbes were unable to consume sufficient $H_{2}$ and $CO_{2}$ to fulfill both their metabolic needs and generate biomass. The second to last data point in Figure \ref{fig:my_label14} shows a scenario where there is signature of life on the planet, but life has not fully gone extinct. In this case the total population of the planet was around 6.5 individuals due to the limitations in representing small populations in code designed for populations of the order of $10^{24}$. In this case the microbes at such small numbers have no measurable impact on their planet and such a scenario could represent the remnants of a collapsed biosphere `clinging on' in small pockets on its planet \citep{wilkinson2007fundamental}. In these experiments we again set the other microbe parameters to the values in Table \ref{table4}.

Again we found that increasing the protein content also only had a slight impact on the level of methane in the atmosphere, while the planet supported a thriving biosphere. As we increased the $CH_{2}O$ content of a microbe cell we found again no measurable change in the temperature of the planet, a significant reduction in the total population for increasing cell protein content, a small reduction in the level of atmospheric methane, and an increase in the average methane output per microbe per hour. These results again show that significantly changing the parameters detailing the population dynamics of the simple biosphere only has a small impact on the level of atmospheric methane resulting from the biospheres' metabolic activity, for biological parameters that allow for a successful biosphere. Thus the underlying population dynamics of the biosphere do not significantly impact the resulting biosignature we would expect to measure for our $H_{2}$ limited biosphere.

\subsection{Burial rate}

We can combine the data from our biological experiments if instead of measuring the biosphere in terms of microbe population, ATP generation or death rate, we measure the biological burial rate of $H_{2}$. When the microbes die, we assume that the dead cells fall to the bottom of the ocean and are buried, and therefore hydrogen is removed via this process as the microbes cells are built from $CH_{2}O$. The burial rate depends on the population of the microbes, the protein content of their cells, and the rate at which microbes are removed from the system. We consider here the burial of rate $H_{2}$ as our biosphere is $H_{2}$ limited. If carbon dioxide instead was the limiting factor on microbe growth, the burial rates of carbon and oxygen would more important processes to study. 

\begin{figure}
    \centering
    \includegraphics[scale=0.53]{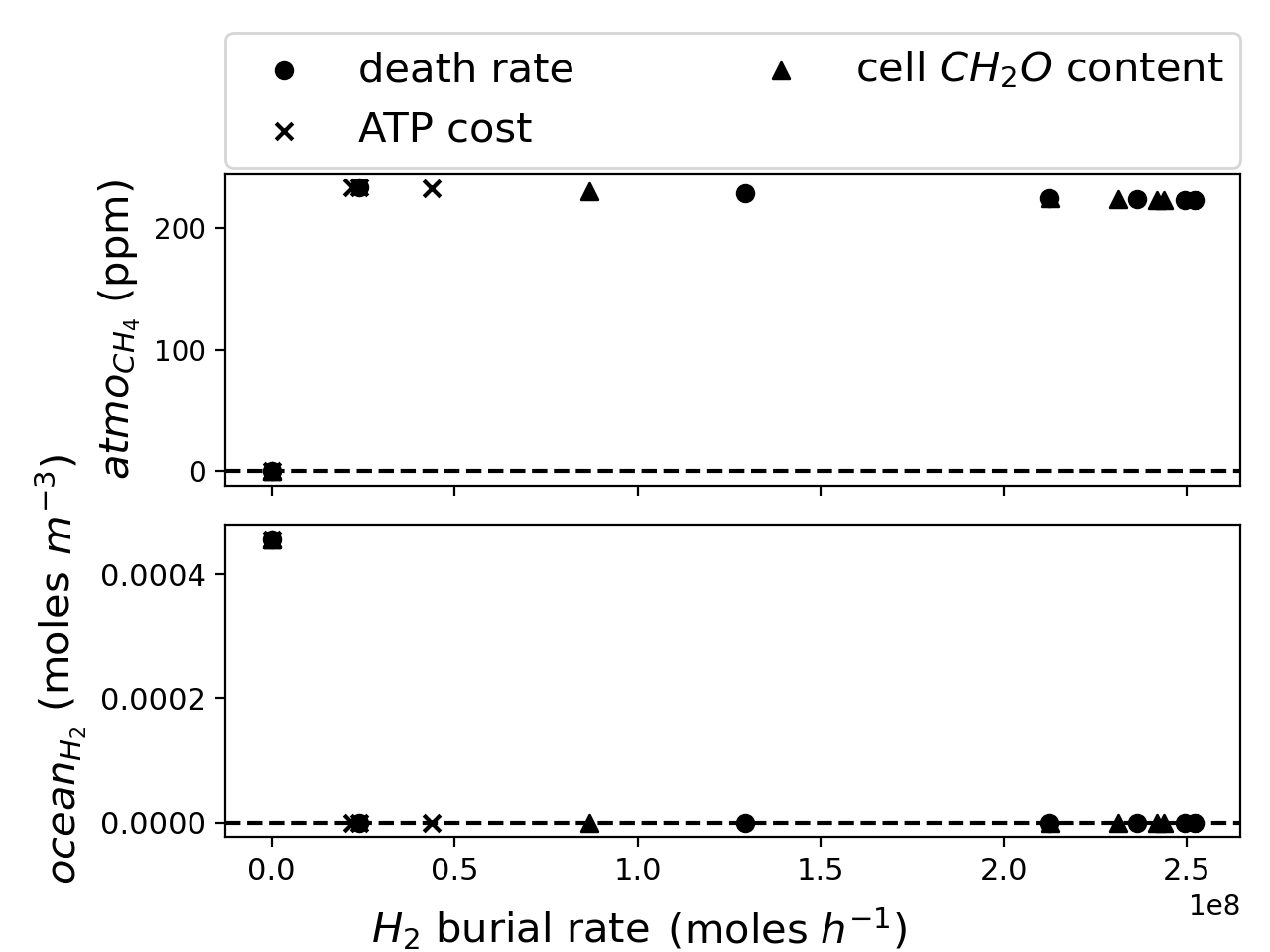}
    \caption{Top panel showing the relationship between $H_{2}$ burial rate and the level of atmospheric methane ($atmo_{CH_{4}}$) for all the data from changing the biological parameters. The dashed line marks $atmo_{CH_{4}} = 0$. Bottom panel showing the relationship between $H_{2}$ burial rate and the concentration of $H_{2}$ in the ocean ($ocean_{H_{2}}$) for all the data from changing the biological parameters. The dashed line marks $ocean_{H_{2}} = 0$.}
    \label{fig:my_label16}
\end{figure}

Figure \ref{fig:my_label16} shows the relationship between $H_{2}$ burial rate and level of methane in the atmosphere in the top panel, and the concentration of $H_{2}$ in the ocean in the bottom panel for all of the data presented thus far from our experiments in Sections \ref{Section:death-rate} and \ref{Section:additionalbiologicalparameters}. We find a negative linear relationship between burial rate and the level of atmospheric methane $atmo_{CH_{4}}$ for experiments where the biosphere avoided extinction or collapse. We also find for these experiments that universally $H_{2}$ in the ocean, $ocean_{H_{2}}$ was drawn down to a concentration of zero. For experiments where extinction or collapse occurred, the data points in Figure \ref{fig:my_label16} are clustered at $atmo_{CH_{4}} = 0$ and $ocean_{H_{2}} \approx 0.00042$. 

Looking at the combined data we find that a large change in the burial rate, which reflects a large change in the population dynamics of our microbes, translates to a much smaller change in the abundance of methane in the atmosphere. The negative linear relationship between $H_{2}$ burial rate and atmospheric methane levels is found as the microbes growth is limited by the availability of $H_{2}$, and so microbes consume all $H_{2}$ in the ocean. This means that all hydrogen in the ocean is used in either building biomass via Equation \ref{Eq:biomass}, or to generate ATP via Equation \ref{Eq:metabolism}. Therefore if more $H_{2}$ is used for ATP generation, less is available for biomass production which scales linearly with biological $H_{2}$ burial rates as this is calculated by multiplying the total biomass of the biosphere by the death rate of the microbes. Note that in both Equations \ref{Eq:metabolism} and \ref{Eq:biomass} $H_{2}O$ is also a product of the reactions. For every 2 moles of $H_{2}$ used to create 1 mole of $CH_{4}$ an additional 2 moles of $H_{2}$ will be use to produce 2 moles of $H_{2}O$ in Equation \ref{Eq:metabolism}, and for every 1 mole of $H_{2}$ converted into 1 mole of biomass ($CH_{2}O$), 1 mole of $H_{2}$ will be converted in to $H_{2}O$ via Equation \ref{Eq:biomass}.

This demonstrates that it is not necessary for a remote observer of our model planet to know the specific ATP requirements or death rates of life on the planet in order to make predictions on the level of atmospheric methane they would expect from a methane producing life form, as long as the combined biological parameters allow for a thriving biosphere on the planet. Instead life in our model can be understood in terms of a process that convert $CO_{2}$ and $H_{2}$ to $CH_{4}$ (as in Equation \ref{Eq:metabolism}) at a rate set by the availability of the limiting resource, in this case $H_{2}$. Therefore we find that in order to make robust biosignature predictions it is more important to understand the abiotic processes occurring on a planet than it is to understand the population dynamics of any alien life.

\subsection{Changing the $H_{2}$ influx}
\label{Section:H2_influx}

\begin{figure}
    \centering
    \includegraphics[scale=0.53]{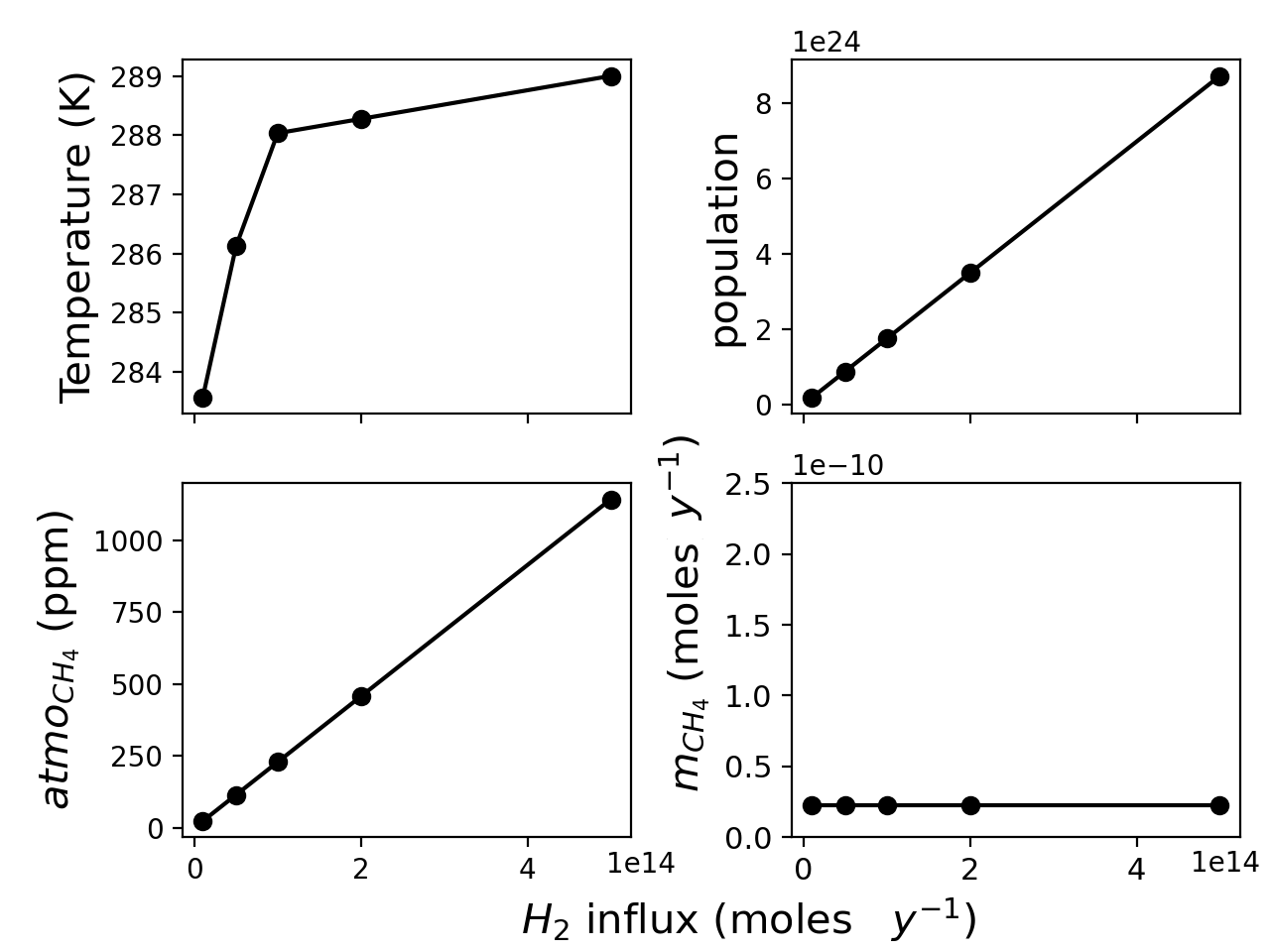}
    \caption{Panels showing the sensitivity of the surface temperature, microbe population, atmospheric $CH_{4}$ content ($atmo_{CH_{4}}$), and the average $CH_{4}$ output per microbe ($m_{CH_{4}}$) to changing the abiotic $H_{2}$ influx to the atmosphere.}
    \label{fig:my_label17}
\end{figure}

In Sections \ref{Section:death-rate} and \ref{Section:additionalbiologicalparameters} we showed that changing the biological parameters only had a small impact on the abundance of methane in the planet's atmosphere. Here we demonstrate that the impact of changing the availability of $H_{2}$, the limiting resource on microbe growth conversely has a large impact in the abundance of atmospheric methane. Figure \ref{fig:my_label17} shows the impact of changing the abiotic influx of $H_{2}$ from 0.1 times the value in Table \ref{table1} up to 5 times this value. In Figure \ref{fig:my_label17} we find that increasing the influx of $H_{2}$ corresponds to a linear increase in the microbe population and a large linear increase in the level of $CH_{4}$ in the atmosphere. The average surface temperature of the planet also significantly increases. However, the average amount of methane produced per microbe per hour, $m_{CH_{4}}$, remains constant.

Comparing Figure \ref{fig:my_label17} with Figure \ref{fig:my_label16} we see that changing the abiotic input of $H_{2}$ to the atmosphere has a much larger impact on the atmospheric methane level than changing any biological parameter. This demonstrates that, where life is limited by a resource, changing the availability of the limiting resource has a much larger impact on the planet than changing biological parameters such as death rate or ATP maintenance cost. 

\begin{figure}
    \centering
    \includegraphics[scale=0.53]{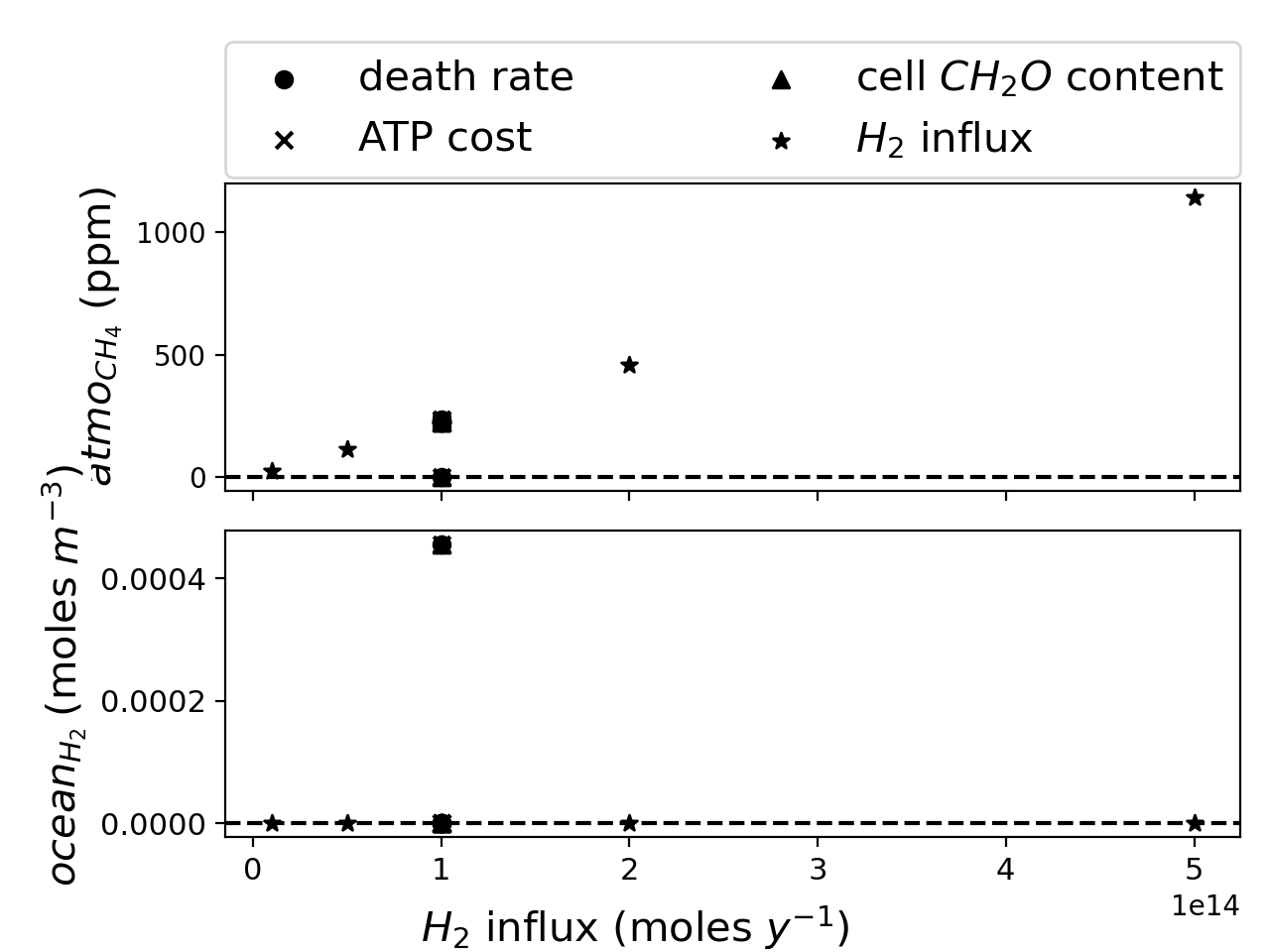}
    \caption{Figure collating all data from changing the biological parameters, and changing the $H_{2}$ influx. Top panel shows atmospheric $CH_{4}$ ($atmo_{CH_{4}}$) vs $H_{2}$ influx. The dashed line marks $atmo_{CH_{4}} = 0$. Bottom panel shows the ocean concentration of $H_{2}$ ($ocean_{H_{2}}$) vs $H_{2}$ influx. The dashed line marks $ocean_{H_{2}} = 0$.}
    \label{fig:my_label20}
\end{figure}

Figure \ref{fig:my_label20} shows the data from all the experiments, both changing biological parameters and changing the $H_{2}$ influx, and shows the level of atmospheric methane against $H_{2}$ influx in the top panel, and the concentration of $H_{2}$ in the ocean against $H_{2}$ influx in the bottom panel. As we kept the $H_{2}$ influx constant for each of our biological experiments these data points are all plotted against the same $H_{2}$ influx. The data points from the biological experiments for planets with a thriving biosphere are clustered very close together towards the lower left hand part of Figure \ref{fig:my_label20} showing the small impact these parameters have on methane abundance in the atmosphere.This contrasts with the strong positive relationship seen between $H_{2}$ influx and atmospheric methane abundance. Again, experiments that resulted in extinction or only sparse life are clustered at $atmo_{CH_{4}} = 0$.

Looking at Figure \ref{fig:my_label20} we can clearly see that the influx of $H_{2}$, and thus the availability of the limiting resource to the biosphere, is more important for determining the level of atmospheric methane we expect as a result of our simple biosphere, than any of the parameters governing the population dynamics of the microbes. Therefore for a remote observer of our model planet hoping to understand the amount of atmospheric methane they might expect to see in the atmosphere for a potential $H_{2}$ limited biosphere dominated by methane producing life, it would be more important for them to accurately model the $H_{2}$ influx in the atmosphere, the rate of hydrogen escape, and the breakdown of methane, rather than focusing on the population dynamics of life on the planet.

It is important to note these relationships only hold true when our model life is able to exploit hydrogen in the ocean to the same extent in each experiment. If the life became limited by some other factor in one or more of these experiments, these relationships between the biological parameters, $H_{2}$ availability and level of methane in the atmosphere would break down. 

As discussed in Section \ref{Section:microbes} real methanogens cannot draw levels of $H_{2}$ down to zero as they are limited by diffusion. For real microbes the limit to which they can draw down their limiting nutrient will depend on factors such as their cell size and shape, and the rate at which they require nutrients to maintain a stable population \citep{beveridge:1988, siefert:1998, koch:1996, schulz:2001, young:2006}. We have assumed here that in each experiment the microbe have evolved to exploit $H_{2}$ to the same minimum limit, taken to be zero for simplicity. These results demonstrate that different biospheres that can exploit $H_{2}$ to the same extent, whatever the underlying population dynamics might be, will result in very similar biosignatures. Determining the limit to which life can exploit its limiting resource will be key to understanding possible biosignatures on exoplanets.

\subsection{Adding more realistic energy harvesting}
\label{Section:thermoresults}

We added complexity to our model by determining the amount of energy obtained from a microbe's metabolism by calculating the free energy change of the chemical reaction under the environmental conditions (see Equation \ref{Equation:thermo}). Therefore instead of obtaining a fixed amount of energy per mole of $CH_{4}$ produced (referred to as energy scheme (a)), the amount of energy now depends on the temperature and on the concentration of $H_{2}$, $CO_{2}$, and $CH_{4}$ in the ocean (referred to as energy scheme (b)).

Another regime is now possible in the model. When the energy obtained from a mole of $H_{2}$ remains fixed, the biosphere either exploits $H_{2}$ reserves in the ocean to zero, or collapses and goes extinct or nearly extinct. When the energy obtained per mole of $H_{2}$ is determined by Equation \ref{Equation:thermo}, as the concentration of $H_{2}$ changes in the ocean, the energy obtained per mole of $H_{2}$ also changes. Therefore a biosphere with a certain biological parameter setup can become limited by the energy obtained from Equation \ref{Equation:thermo} preventing it from drawing down the ocean concentration to zero.

\begin{figure}
    \centering
    \includegraphics[scale=0.48]{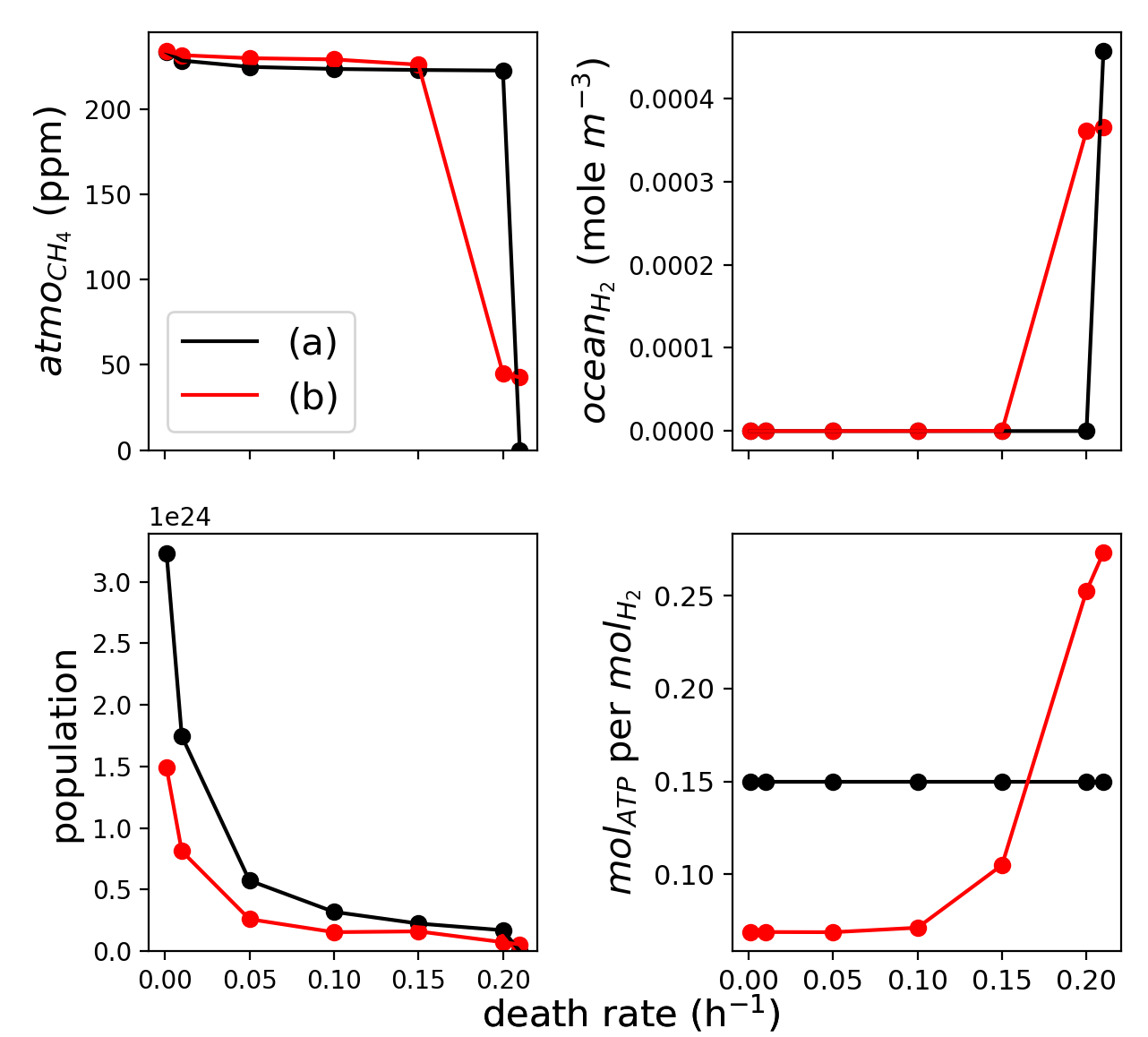}
    \caption{Panels showing the sensitivity of the atmospheric level of methane ($atmo_{CH_{4}}$), the ocean concentration of hydrogen ($ocean_{H_{2}}$), the microbe population and the moles of ATP generated per mole of $H_{2}$ consumed ($mol_{ATP}$ per $mol_{H_{2}}$) to changing death rate for the two different energy schemes: (a) where microbes obtain a fixed amount of energy per mole of $CH_{4}$ produced, and (b) where this energy is given by the free energy change of $CO_{2} + 4H_{2} \rightarrow CH_{4} + 2H_{2}O$ - the microbes' metabolism.}
    \label{fig:thermo_deathrate}
\end{figure}

Figure \ref{fig:thermo_deathrate} shows the sensitivity experiments changing the death rate of the microbes for both energy scheme (a) shown in black and (b) shown in red. The panels in figure \ref{fig:thermo_deathrate} show how the atmospheric level of methane ($atmo_{CH_{4}}$), the ocean concentration of hydrogen ($ocean_{H_{2}}$), the microbe population and the moles of ATP generated per mole of $H_{2}$ consumed ($mol_{ATP}$ per $mol_{H_{2}}$) change with death rate for the two different energy schemes.

The first panel in Figure \ref{fig:thermo_deathrate} shows that for smaller death rates, the two energy schemes yield very similar levels of atmospheric methane. However the two schemes diverge above a death rate of $15\%$. For a death rate of $20\%$, the biosphere under energy scheme (a) is $H_{2}$ limited and produces a similar level of atmospheric methane as biospheres with smaller death rates. The biosphere with a death rate of $20\%$ under energy scheme (b) however does not draw the level of $H_{2}$ in the ocean to zero. Instead the $H_{2}$ concentration in the ocean remains higher, and the corresponding level of atmospheric methane is greatly reduced. With a high death rate, microbes must reproduce faster to maintain a stable population. When the energy obtained per mole of $H_{2}$ is fixed, the microbe is limited only by hydrogen availability and $C_{max}^{H_{2}}$ - the maximum rate at which it can consume $H_{2}$. When however the energy obtained per mole of $H_{2}$ is determined by Equation \ref{fig:thermo_deathrate}, the energy obtained per mole of $H_{2}$ consumed will change as the concentration of $H_{2}$ in the ocean changes. Therefore microbes will be able to draw the level of hydrogen down in the ocean to the level where they obtain sufficient energy from $C_{max}^{H_{2}}$ moles of $H_{2}$ to maintain a stable population.

\begin{figure}
    \centering
    \includegraphics[scale=0.35]{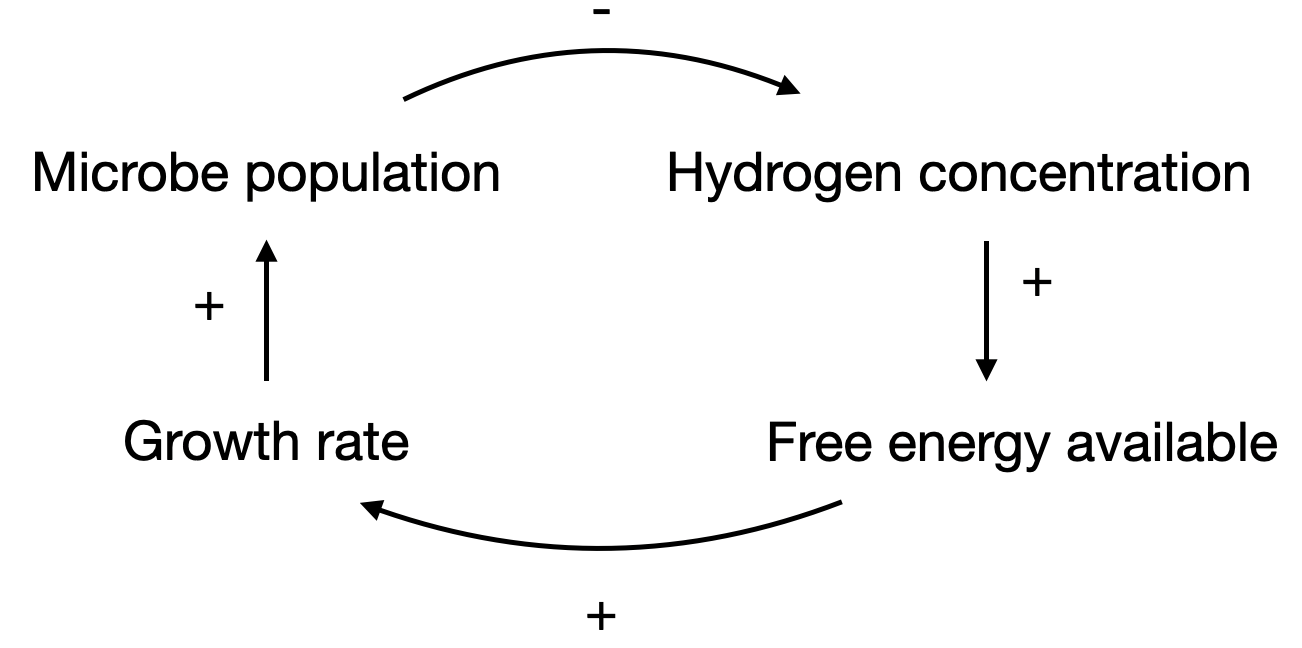}
    \caption{Diagram showing the feedback loop between the $H_{2}$ ocean concentration, free energy available, growth rate of the microbes and the total population of the biosphere. A + sign indicates than an increase in the source leads to an increase in the sink. A - sign indicates that an increase in the source leads to a decrease in the sink.}
    \label{fig:feedback2}
\end{figure}

A slightly different feedback loop comes into effect when the microbe biosphere is limited by the energy yield of Equation \ref{Equation:thermo}. Now the free energy available plays a role in determining the microbe population the planet can support for any biological parameter combination that pushes the system out of the purely $H_{2}$ availability limited regime e.g. a high death rate of $20\%$ as shown in Figure \ref{fig:thermo_deathrate}. Figure \ref{fig:feedback2} shows the stabilising feedback loop between the $H_{2}$ ocean concentration, free energy available, growth rate of the microbes and the total population of the biosphere.

\begin{figure}
    \centering
    \includegraphics[scale=0.48]{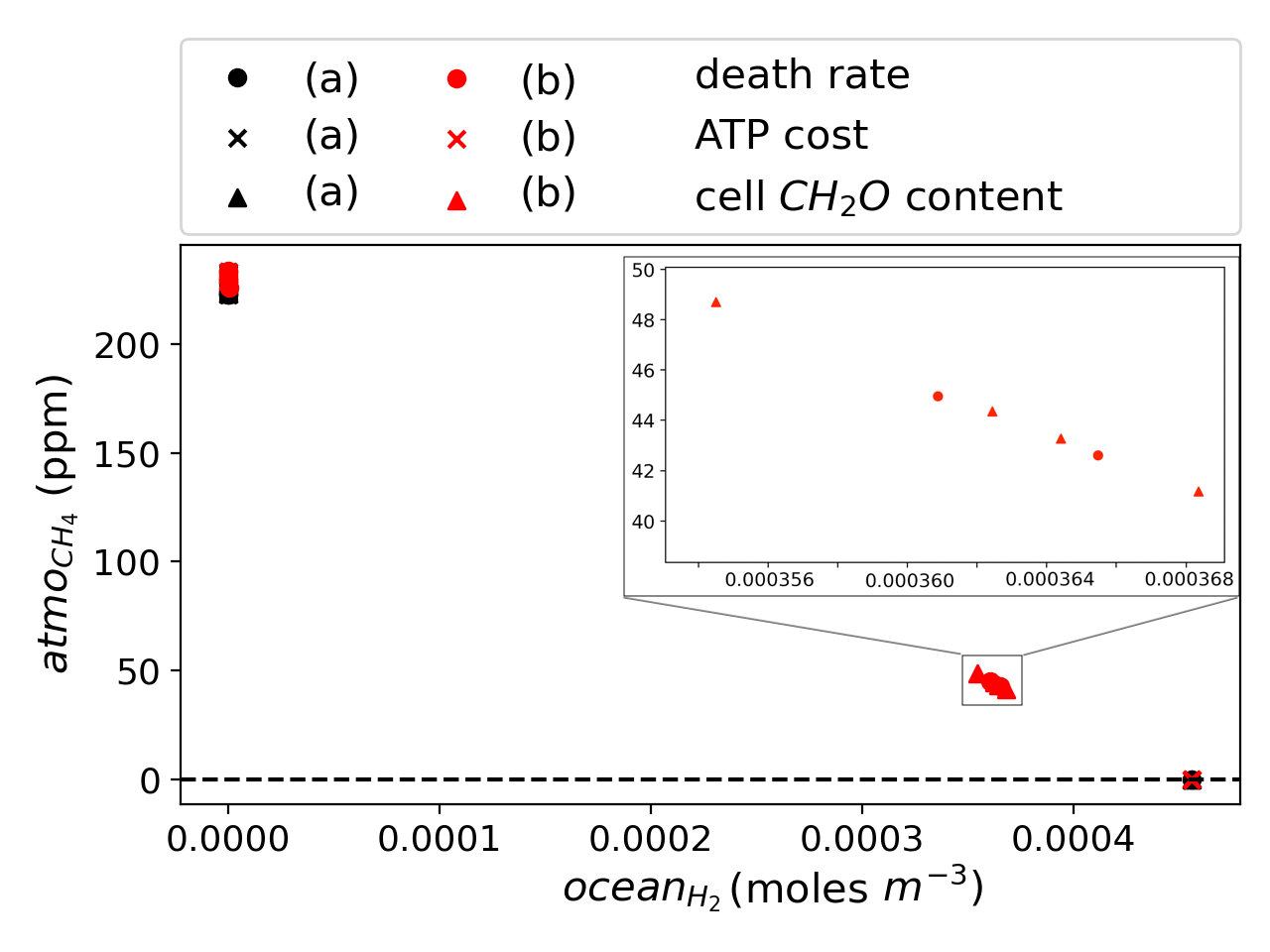}
    \caption{Atmospheric methane abundance ($atmo_{CH_{4}}$) vs ocean $H_{2}$ content ($ocean_{H_{2}}$) for experiments changing biological parameters of death rate, ATP maintenance cost, and protein cell content for experiments where (a) microbes obtain a fixed amount of energy per mole of $CH_{4}$ produced, and (b) where this energy is given by the free energy change of $CO_{2} + 4H_{2} \rightarrow CH_{4} + 2H_{2}O$ - the microbes' metabolism. The dashed line marks $atmo_{CH_{4}} = 0$.}
    \label{fig:my_labelyyy}
\end{figure}

We repeated the previous experiments of changing the biological parameters while utilising energy scheme (b).  Figure \ref{fig:my_labelyyy} shows the abundance of atmospheric methane against the ocean concentration of $H_{2}$ for all the biological experiments from Sections \ref{Section:death-rate} and  \ref{Section:additionalbiologicalparameters}, labeled (a) and shown in black, combined with results from repeating these experiments but with realistic energetics from Equation \ref{Equation:thermo}, labeled (b) and shown in red. We see three clear groupings of data points in Figure \ref{fig:my_labelyyy} - a cluster of points for an ocean concentration of $H_{2}$, where the corresponding levels of atmospheric $CH_{4}$ - $atmo_{CH_{4}}$, are clustered around roughly $atmo_{CH_{4}} = 220$. This is the regime where the biosphere is able to exploit all the $H_{2}$ in the ocean and is solely limited in growth by the availability of $H_{2}$. The total populations of the different biospheres in these cases leads to only slight changes in the abundance of atmospheric methane.

Another cluster of points are seen in Figure \ref{fig:my_labelyyy} between $ocean_{H_{2}} = 0.003$ and $ocean_{H_{2}} = 0.004$. Here we find only data points from experiments under energy scheme (b) and this is the regime where the biosphere is limited by the free energy available from Equation \ref{Equation:thermo}. For biospheres requiring more energy to maintain a stable population, a higher concentration of $H_{2}$ in the ocean is necessary, as per the feedback loop in Figure \ref{fig:feedback2}. In this cluster of data points an inverse linear relationship exists between $atmo_{CH_{4}}$ and $ocean_{H_{2}}$ - as the level of $H_{2}$ in the ocean increases, the level of methane in the atmosphere decreases linearly. Therefore determining the extent to which a biosphere can exploit $H_{2}$ in the ocean determines the biosignature.

Figure \ref{fig:my_labelyyy} shows a third cluster of data points where $atmo_{CH_{4}} = 0$. These are experiments where life went extinct and so no biosignature is present on the planet. Figure \ref{fig:my_labelyyy} shows a clear relationship between $H_{2}$ availability in the ocean and $CH_{4}$ levels in the atmosphere for all experiments.

\begin{figure}
    \centering
    \includegraphics[scale=0.48]{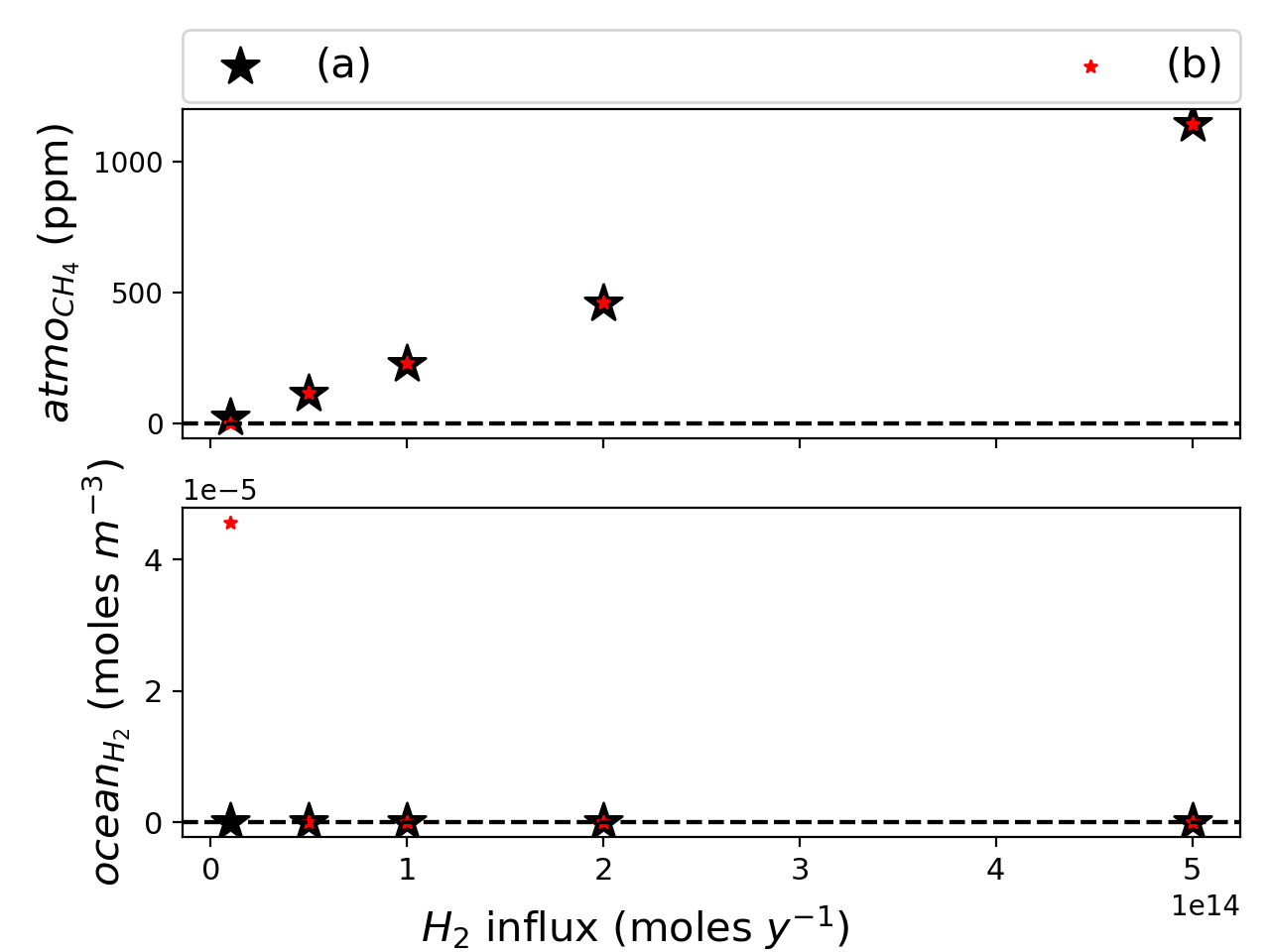}
    \caption{Atmospheric methane abundance ($atmo_{CH_{4}}$), and ocean concentration of $H_{2}$ ($ocean_{H_{2}}$) vs $H_{2}$ influx for all experiments, both changing biological parameters, and changing the $H_{2}$ influx for two scenarios where (a) microbes obtain a fixed amount of energy per mole of $CH_{4}$ produced, and (b) where this energy is given by the free energy change of $CO_{2} + 4H_{2} \rightarrow CH_{4} + 2H_{2}O$ - the microbes metabolism. The dashed lines mark $atmo_{CH_{4}} = 0$ and $ocean_{H_{2}} = 0$ in the top and bottom panels respectively.}
    \label{fig:my_label22}
\end{figure}

Figure \ref{fig:my_label22} shows the data for changing $H_{2}$ influx vs methane content when adopting energy scheme (b) (shown in red), and the corresponding results from the previous experiments (from Section \ref{Section:H2_influx}) where energy scheme (a) was used (shown in black). We find again that changing the $H_{2}$ influx has a significant impact of the level of methane in the atmosphere and these data points very closely overlap. We see a very slight difference between the two energy schemes in the data points for the lowest $H_{2}$ influx, but in this case, in the experiment using energy type (b) life went extinct on the planet resulting in no atmospheric methane.

Our model setup has included only one species of life and so there is no inter-species competition for resources. Where the biosphere is able to exploit all the $H_{2}$ in the ocean, the exact details of the microbes, e.g. death rate or ATP maintenance cost, have only a minimal impact on the abundance of atmospheric $CH_{4}$. For two species with differing biological parameters but the same ability to exploit $H_{2}$ in the ocean, neither will have a selective advantage over the other, and the methane biosignature is only minimally impacted by whichever species becomes dominant. For situations where the biosphere becomes limited by the free energy available from Equation \ref{Equation:thermo}, these parameters can now have an impact on the limit to which the biosphere can draw $H_{2}$ down in the ocean, and this will then impact the resulting abundance of atmospheric methane. However in a scenario with multiple species, the species capable of drawing $H_{2}$ down to the lowest concentration in the ocean would have the selective advantage as its growth rate will not become limited as quickly as other species as the $H_{2}$ concentration in the ocean drops. Therefore we would expect the species that can draw down ocean $H_{2}$ to the lowest concentration to out-compete other species \citep{tilman2020resource}. Determining the methane biosignature for a nutrient limited biosphere would therefore depend on determining the theoretical limit to which life can evolve to exploit the limiting nutrient.

\subsection{Biomass as a metric for predicting the methane biosignature}

Methods of predicting potential biosignatures include a biomass-based model to estimate the plausibility of exoplanet biosignature gases developed by \cite{Seager2013}. In this biomass-based model potential biosignatures are calculated based on Earth-based measurements of maximum biomass per area, and maximum gas output rates for different species. \cite{Seager2013} combine this data to obtain a theoretical maximum biosignature strength to be used to verify future possible biosignature observations. The biomass-based model determines whether the abundance of a proposed biosignature gas in the atmosphere of an exoplanet would be biologically viable to be the product of life. The work in this paper presents a different approach for predicting biosignatures by instead focusing on determining the possible limiting factors on a biospheres growth to constrain possible biosignatures. In this approach it then becomes crucial to understand the availability of the limiting resource which in the case of an $H_{2}$ limited biosphere as considered in this work, will depend factors such as the level of volcanic activity, and the rate of $H_{2}$ loss to space.

\begin{figure}
    \centering
    \includegraphics[scale=0.48]{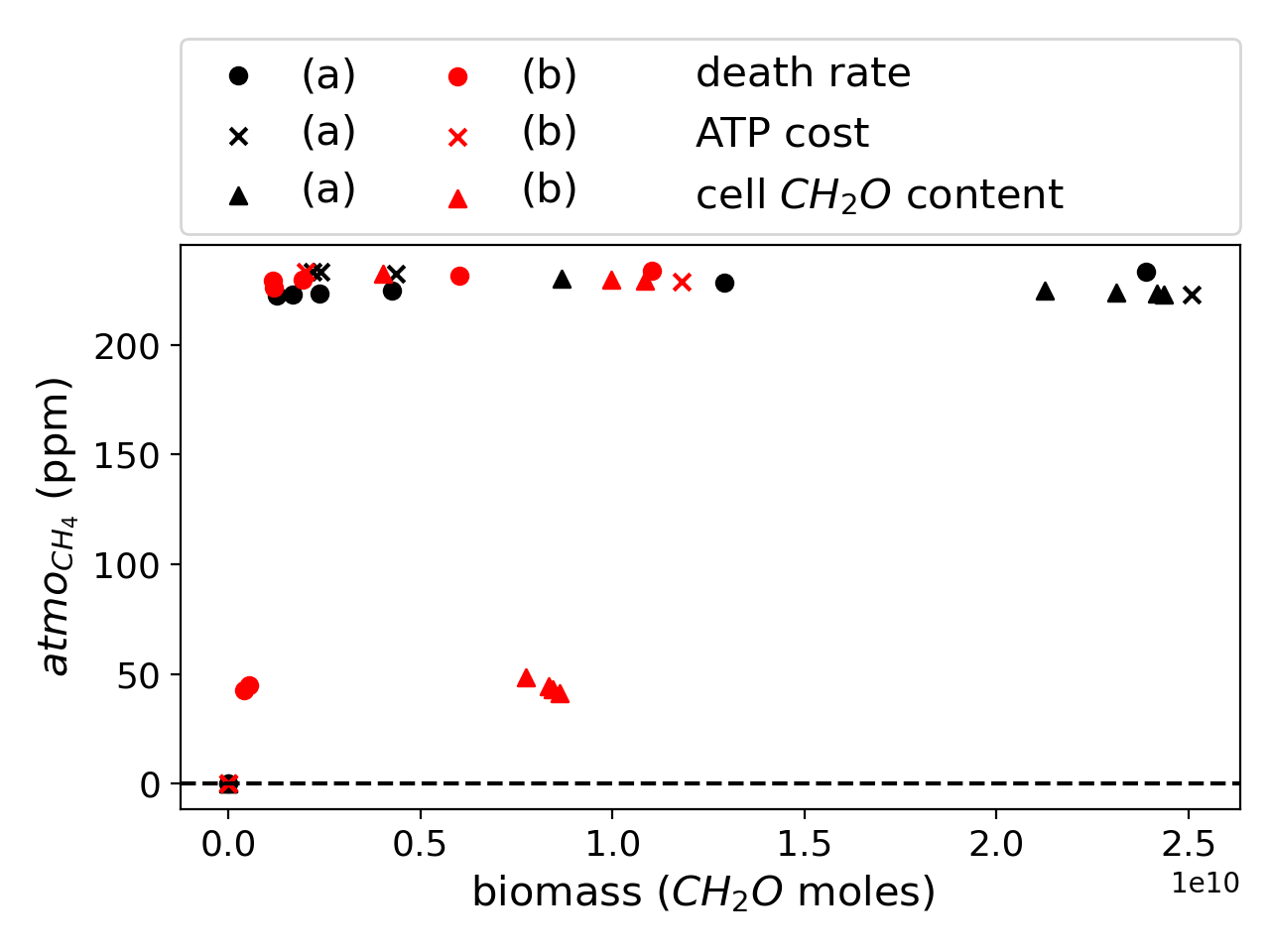}
    \caption{Atmospheric methane abundance ($atmo_{CH_{4}}$) vs biomass (moles of $CH_{2}O$ contained within the biosphere) for experiments changing biological parameters of death rate, ATP maintenance cost, and protein cell content for experiments where (a) microbes obtain a fixed amount of energy per mole of $CH_{4}$ produced, and (b) where this energy is given by the free energy change of $CO_{2} + 4H_{2} \rightarrow CH_{4} + 2H_{2}O$ - the microbes' metabolism. The dashed line marks $atmo_{CH_{4}} = 0$.}
    \label{fig:my_labelxxx}
\end{figure}

Figure \ref{fig:my_labelxxx} shows the data from our experiments in changing the biological parameters for both energy schemes (Sections \ref{Section:death-rate} and \ref{Section:additionalbiologicalparameters}) and shows how the level of atmospheric methane $atmo_{CH_{4}}$ changes with the biomass of the biosphere. The biomass is calculated in terms of number of moles of $CH_{2}O$ contained within the living microbe cells. The three regimes: $H_{2}$ availability limited, energetically limited, and biosphere collapse/extinction are clearly seen in Figure \ref{fig:my_labelxxx} separated along the y-axis however no clear relationship emerges between biomass and $atmo_{CH_{4}}$. Figure \ref{fig:my_labelxxx} shows that large variations in biomass can result in very similar levels of $CH_{4}$ and so in the case of a nutrient limited biosphere biomass would be a poorer metric for predicting a possible type I biosignature than determining the limiting nutrient of the biosphere and the biosphere's ability to exploit this nutrient.

\section{Summary}
\label{Section:summary}

We have shown that for a simple single-species $H_{2}$-limited biosphere, the underlying population dynamics of the microbes are largely irrelevant to the abundance of atmospheric methane in the atmosphere - taken in these experiments as our biosignature - instead the limit to which the biosphere can exploit $H_{2}$ in the ocean is more fundamental. While different biospheres might have different total populations, if two differing biospheres are capable of drawing down ocean concentrations of $H_{2}$ to the same limit, their relative population dynamics only minimally impact the resulting biosignature. We found that the influx of $H_{2}$ to the atmosphere of the planet had a far stronger impact on the level of $CH_{4}$ in the atmosphere. Understanding the abiotic processes governing the availability of the limiting resource to the biosphere, in this case $H_{2}$, and the biosignature gas, $CH_{4}$, are crucial for biosignature predictions for nutrient limited biospheres. The highly simplified abiotic environment of our model planet allows the impact on the atmospheric composition due to the population dynamics of the biosphere to be clearly determined, however for forming biosignature predictions for real planets, a far more sophisticated representation of the abiotic environment will be necessary. 

We have considered a type I biosignature in this model as classified in \cite{Seager2013} where the biosignature is generated as a by-product from microbial energy extraction and the results from this work are only applicable to forming predictions for biosignatures of this type. Biosignatures resulting from processes other than energy extraction of microbes might behave differently with the biological parameters explored in this work.

We found that adding complexity to this model in the form of an accurate representation of the energy obtained from the microbes' metabolisms based on the change of free energy of converting $H_{2}$ and $CO_{2}$ to $CH_{4}$ and $H_{2}O$ does not change these results while the biosphere draws $H_{2}$ in the ocean to zero. As the microbe parameters are increased to the limit where a biosphere with energy scheme (a) would goes extinct, an energetically limited regime emerges where a viable biosphere is still possible but now only for higher concentrations of $H_{2}$ in the ocean. As the energy obtained per moles of $H_{2}$ consumes depends on the ocean concentrations of $CO_{2}$, $H_{2}$ and $CH_{4}$ under energy scheme (b), a more gradual decline in the biosignature occurs, where the level of $CH_{4}$ in the atmosphere decreases linearly as the concentration of $H_{2}$ increases in the ocean. In this work we have modeled biospheres consisting of only a single species of microbe and so inter-species competition is not present. However if multiple species limited by $H_{2}$ coexisted, the species capable of drawing $H_{2}$ down to the lowest concentration would have the selective advantage and out compete the others \citep{tilman2020resource}. Therefore, inter-species competition would lead to limiting nutrients being drawn down to the lowest limit possible. Determining this lowest possible limit for each potential limiting nutrient will be key for making biosignature predictions using this method.

These results help deepen our understanding of life-planet interactions. It reduces the need to make unnecessary assumptions about alien life based on life on Earth. These results show that when considering a nutrient-limited biosphere it is more important to accurately model the processes that regulate the availability of the limiting nutrient, and determine the limit to which life can exploit this nutrient, than it is to model any specific population dynamics for that biosphere. In our model example these key processes are: the influx of $H_{2}$ to the atmosphere, the rate of $H_{2}$ loss to space, and the rate of methane breakdown back to $H_{2}$ and $CO_{2}$. Changing the rate of any of these processes will have a much stronger impact on the abundance of methane we expect to find in the atmosphere of our model planet than changing any parameter dealing with the population dynamics of the microbes. This understanding is already used in studies of Earth history to recreate past climates \citep{herman:2005, Kharecha:2005, bruggeman:2014, Lenton:2018, zakem:2020}, and the work in this study is supported by models of Earth's biosphere. 

Identifying possible metabolic pathways will of course be key to understanding potential biosignatures, as we will need to know what byproducts we expect from various types of life on any potentially inhabited planet. Quantifying the free energy available to a biosphere will also be important, as will estimating recycling efficiency once multiple types of life are involved. However, understanding detailed population dynamics of a potential alien biosphere will not be necessary for us to predict potential type I biosignatures for nutrient limited biospheres. Given that we cannot go and measure any potential alien biology in a lab, and that our understanding of life is inherently biased towards Earth-based life, this significantly reduces the number of assumptions needed to accurately model a proposed alien biosphere under nutrient limitation, and helps us avoid biases based on our understanding of specific organisms on Earth.

This method for predicting possible type I biosignatures depends on understanding the abiotic sources of nutrients available to a biosphere. However, remote detection of, for example, levels of volcanic activity are unfeasible, and predicting these factors for any planet will depend on modeling and further developing our understanding of planet formation. As so many unknowns exist and will continue to exist for any potential future biosignature detections, multiple methods for biosignature verification are required to increase confidence that a potential biosignature is actually due to life. We hope that this work provides an additional tool to the astrobiology community to help verify any possible future biosignature detections.

\section{Next steps}
\label{Section:nextsteps}

The results presented here are highly simplified to enable us to pull out clear relationships between various parameters and the strength of life's impact on its host planet when limited by nutrient availability. We have only considered a planet with a single life form, however any life-bearing exoplanet is more likely to have a diverse biosphere. A diversity of metabolisms are deeply rooted in the `tree' of life on Earth, and the earliest fossil evidence for life on Earth is of 5 morphologically distinct species of microbe indicating that diversification happened quickly \citep{Schopf:2018}. Therefore exploring more complex biospheres will be a necessary step in increasing the applicability of this approach to potentially inhabited exoplanets. Additionally, abstract models of ecology predict that higher diversity ecosystems will on average persist for longer than low diversity ecosystems \citep{CHRISTENSEN:2002, ARTHUR:2022}. This would make it statistically more likely to observe biosignatures produced by complex ecosystems rather than simple ones increasing the importance of modeling these scenarios. 

The results in this paper only hold true where our microbe life is limited by a chemical resource, in this case $H_{2}$. When this is no longer true and some other factor is limiting the biosphere this new limiting factor will then become the key parameter to understanding how potential biosignatures might manifest on a planet. The atmospheric chemistry in our model was kept very simple and for future work we intend to move away from this simple framework to more sophisticated models of atmospheric chemistry and more realistic models of biogeochemistry, adapted from modeling Earth's history \citep{Daines:2017, Lenton:2018}. These results show that when considering a nutrient limited life-form we can reduce the bulk of the impact on its host planet down to its metabolism, and the availability of the limiting nutrient. This study takes a step towards allowing us to insert a very simple biological framework into more sophisticated climate models to achieve robust predictions on possible biosignatures. As discussed, determining the limit to which any lifeform can exploit its limiting nutrient will be key for accurate biosignatures predictions.

Of course not all life is nutrient limited, on Earth much of our biosphere is photon limited, and in future work we aim to recreate the experiments demonstrated here but for a photon limited life-form to determine the minimal assumptions needed about life existing under those circumstances. This would allow us to model both life limited by nutrients, and by photon availability in a simple robust manner allowing us to form hypotheses for potential biosignatures for either case.

We also hope to use this understanding to consider possible biosignatures on `Super Earths' - planets that have a radius of 1.25 - 2 times that of Earth's \cite{Fressin:2013}. Super Earths are some of the most commonly found planets with current observational limitations which makes them interesting candidates in the search for biosignatures. Some of these planets are theorised to have atmospheres far richer in hydrogen than Earth's due to less hydrogen loss to space \citep{Seager:2013b}. Our test model setup above explores scenarios where life is limited by hydrogen. On a hydrogen rich super Earth this may no longer be the case. A high concentration of $H_{2}$ could become an important greenhouse gas \citep{Pierrehumbert:2011} providing significant warming to the planet (which is not the case on Earth). How life would balance the requirement of atmospheric $H_{2}$ to keep its planet warm against its need to consume $H_{2}$ would be key for understanding the potential biosignatures that might be possible on such a planet. The temperature dependence of microbe' metabolisms (as the temperature decreases metabolic activity tends to decrease \citep{clarke:2004}) could become an important factor to consider when predicting possible biosignatures on such planets. 

We have focused on methanogens, but can easily incorporate different metabolisms, or scenarios with multiple competing metabolisms. This work represents the first steps in trying to frame our understanding of how a relatively arbitrary life-form may interact with its planetary atmosphere, and to determine the key parameters or factors necessary to explore and characterise this interaction. It is clear that there is much work to do, requiring a large and diverse community, to enable us to be in a position to confidently, and robustly, determine the presence of a biosignature in addition to developing more sensitive and accurate instrumentation. 

\section{Acknowledgments}
\label{Section:acknowledgements}

We would like to thank the reviewer for their helpful feedback on earlier drafts of this paper. Material produced using Met Office Software. We acknowledge use of the Monsoon2 system, a collaborative facility supplied under the Joint Weather and Climate Research Programme, a strategic partnership between the Met Office and the Natural Environment Research Council. This work was supported by a Leverhulme Trust research project grant [RPG-2020-82], a Science and Technology Facilities Council Consolidated Grant [ST/R000395/1], a UKRI Future Leaders Fellowship [MR/T040866/1], and a John Templeton Foundation grant. JE-N would like to thank the Hill Family Scholarship, generously supported by University of Exeter alumnus, and president of the University’s US Foundation Graham Hill (Economic $\&$ Political Development, 1992) and other donors to the US Foundation.

\section{Data Availability Statement}

The code used to generated the data in this study can be found at: \url{https://github.com/nicholsonae/archean\_world}

\bibliographystyle{mnras}
\bibliography{main.bib}

\label{lastpage}
\end{document}